\documentclass[twocolumn,aps,superscriptaddress,longbibliography]{revtex4-1}
\usepackage{graphicx}
\usepackage{dcolumn}
\usepackage{romannum}
\usepackage{simplewick}
\usepackage{bm}
\usepackage{hyperref}
\usepackage[usenames,dvipsnames]{color}
\usepackage{amsmath}
\usepackage{amssymb}
\usepackage{times}

\def\red{|\!|}

\def\pt{\mathbf{T}^{(1)}}

\def\ps{\mathbf{S}^{(1)}}
\def\pso{\mathbf{S}^{(1)}_1}
\def\pst{\mathbf{S}^{(1)}_2}
\def\helec{{\mathbf{H}}_{\rm elec}^{\rm NSD}}

\begin{document}

\title{Fock-space perturbed relativistic coupled-cluster calculations 
of electric dipole polarizability and nuclear spin-dependent parity 
non-conservation in Cs}

\author{Suraj Pandey}
\affiliation{Department of Physics, Indian Institute of Technology,
             Hauz Khas, New Delhi 110016, India}


\author{Ravi Kumar}
\affiliation{Department of Chemistry, University of Zurich,
             Switzerland}

\author{D. Angom}
\affiliation{Department of Physics, Manipur University,
             Canchipur 795003, Manipur, India}

\author{B. K. Mani}
\email{bkmani@physics.iitd.ac.in}
\affiliation{Department of Physics, Indian Institute of Technology,
             Hauz Khas, New Delhi 110016, India}

\begin{abstract}

We implement the Fock-space perturbed relativistic coupled-cluster theory 
to compute the electric dipole polarizability of ground and low lying excited 
states, and nuclear spin-dependent parity violating (NSD-PNC) transition 
amplitudes in Cs. 
Moreover, to check the accuracy of the wavefunctions used in the calculations, 
we compute the excitation energies, E1 transition amplitudes and magnetic dipole 
hyperfine constants for ground and low lying excited states. To improve the 
accuracy of the computed properties, we have incorporated the corrections 
from the relativistic and QED effects in our calculations. The contributions 
from triple excitations are accounted perturbatively. Our results on excitation 
energies, E1 transition amplitudes and hyperfine constants are in good agreement 
with the available experimental results. Our polarizability results using FS-PRCC
theory match well with the experimental values. The values of parity-violating 
transition amplitudes from our calculations are, in general, on the lower side 
of the previous values. From the detail analysis of electron correlations, we 
find that the corrections from the Breit interaction and QED effects are 
important to get accurate results of NSD-PNC amplitudes in Cs. 
The largest cumulative contribution from the Breit and QED corrections
is found to be $\approx$ 3.2\% of the total value. The upper bound on 
the theoretical uncertainty in our calculated NSD-PNC amplitudes is 
estimated to be about 1\%.

\end{abstract}
\pacs{31.15.bw, 11.30.Er, 31.15.am}


\maketitle

\section{Introduction}

The parity nonconservation (PNC) in atoms and ions is one of the
most important phenomena in fundamental physics to understand physics beyond
the Standard Model of particle physics \cite{roberts-15, safronova-18}.
The PNC in atoms manifest in two forms, nuclear spin-independent (NSI) and
nuclear spin-dependent (NSD).  The NSI-PNC is well studied both
experimentally \cite{tsigutkin-09, barkov-79,wood-97} as well as
theoretically \cite{porsev-10, kozlov-01, kozlov-01b, dzuba-12} in several
atoms. To date, the most precise PNC measurement was carried out using Cs atom
at Colorado by Wood {\em et. al.} \cite{wood-97} with an accuracy of 0.35\%.
The most precise theoretical results also correspond to Cs, where
the reported NSI-PNC amplitude is calculated with a theoretical
accuracy of 0.27\% \cite{porsev-09, porsev-10}. Unlike NSI-PNC, the NSD-PNC
is not well explored. The same Cs experiment \cite{wood-97} also observed a
signature of NSD-PNC but the effect was very weak due to strong
NSI-PNC component. The experimental error in the reported NSD-PNC observable
was $\approx 15$\% \cite{wood-97}. In theoretical studies too, it is
challenging to compute NSD-PNC transition amplitudes accurately as it involves
coupling of nuclear and electronic wavefunctions. It is, however, important to
explore NSD-PNC in atomic systems as it is essential to deduce nuclear
properties like the anapole moment (NAM). The NAM, first suggested by
 Zeldovich \cite{zeldovich-58}, is a parity odd nuclear electromagnetic moment
and arises from parity violating phenomena within the nucleus. It should be
noted that, NAM is one of the dominant sources to NSD-PNC. The other two
sources are, the combination of hyperfine and nuclear spin-independent PNC
and spin-dependent Z-exchange between electrons and nucleus.

As theoretical results are essential to obtain NAM from the experimental
data, it is crucial to employ reliable and accurate quantum many-body theory
in the calculations.
The theory must account for electron correlations and relativistic effects to
the highest possible level of accuracy. Among the previous theoretical works
on calculations of NSD-PNC transition amplitudes in Cs, there are two 
calculations which use methods significantly different from the present work. 
The first one is by Johnson and collaborators \cite{johnson-03}, they used
random-phase-approximation (RPA). And, the second one by Safronova and
collaborators \cite{safronova-09a} uses all-order method.
There are two other works \cite{mani-11cs, chakraborty-24} using methods
similar to the present work. Although the two calculations employ similar
coupled-cluster method, there is a difference in the reported values of
the transition amplitudes. An important point to be highlighted is, none of
the previous works incorporate corrections from the Breit interaction, QED
effects and triples excitations in coupled-cluster. It should, however, be
noted that these corrections are important to get accurate and reliable values
of parity violating transition amplitudes. It can thus be surmised that
the availability of accurate theoretical data on NSD-PNC in Cs is a research
gap which needs to be addressed. Addressing this gap is one of the main aims
of the present work

In this work, we implement a Fock-space perturbed relativistic coupled-cluster
(FS-PRCC) theory to compute NSD-PNC electric dipole transition amplitude,
$E1^{\rm NSD}_{\rm PNC}$, for all allowed hyperfine transitions corresponding
to the $6s_{1/2}\rightarrow7s_{1/2}$ transition in Cs. It is to be noted that,
relativistic coupled-cluster (RCC) theory is one of the most reliable
many-body theories for atomic structure calculations. It accounts for
the electron correlations to all-orders of residual Coulomb interaction, and
has been employed to study a plethora of properties in atomic systems.
The implementation of RCC theory including properties calculations of
closed-shell and one-valence atomic systems without perturbation is reported
in our previous work \cite{mani-17}. For properties calculation in the
presence of external perturbation, we had reported the development of 
PRCC theory for closed-shell and one-valence systems in our previous
works \cite{chattopadhyay-12, chattopadhyay-14, 
chattopadhyay-15, ravi-20, ravi-22}. A PRCC theory for PNC calculation
is reported in our recent work  \cite{mani-24}. 
One of the key merits of PRCC in properties calculation is that it does 
not employ the sum-over-states approach \cite{safronova-99,derevianko-99} 
to incorporate the effects of perturbation. In PRCC, summation over 
all the possible intermediate states is subsumed through the 
perturbed cluster operators.

Furthermore, to check the accuracy of the wavefunctions in the small and large
$r$ regions, we calculate the hyperfine constants and E1 transition
amplitudes, respectively. In addition, to check and validate our implementation
of PRCC theory for NSD-PNC, we compute the dipole polarizability, $\alpha$,
of ground state $6s_{1/2}$ and PNC-transition state $7s_{1/2}$ using PRCC.
Except for the angular factor, as the coupled-cluster diagrams for PNC are
similar to the polarizability diagrams obtaining accurate $\alpha$ using
PRCC indicates possibility of obtaining accurate $E1^{\rm NSD}_{\rm PNC}$. To
verify the results, we also use the sum-over-states approach to compute
$\alpha$ for $6s_{1/2}$, $6p_{1/2}$, $7s_{1/2}$ and $7p_{1/2}$ states.
The other key highlight of the present work is, to further improve the accuracy
of our results, we have incorporated the corrections from the Breit interaction,
QED effects and perturbative triples in the calculation.

The remainder of the paper is divided into five sections. In Sec. II, we
provide a brief discussion on the FSRCC and FS-PRCC theories for one-valence
systems. In the same section, we also elaborate the calculation of
$E1^{\rm NSD}_{\rm PNC}$ using FS-PRCC theory and discuss the corresponding
Goldstone diagrams. Then, the coupling of nuclear spin with the electronic
states is also discussed in the same section. In Sec. III, we present and
discuss our results of excitation energy, E1 transition amplitudes and
hyperfine constants, dipole polarizability, and $E1^{\rm NSD}_{\rm PNC}$ in
different subsections. The theoretical uncertainty is presented in Sec. IV of
the paper. Unless stated otherwise, all the results and equations presented
in the paper are in atomic units ($\hbar = m_e = e = 1/{4\pi\epsilon_0} = 1$).

\section{Theoretical methods}
\label{method}

\subsection{FSRCC theory for one-valence systems}

To obtain accurate results in the structure or property calculations of an 
atom or ion, it is essential to use accurate wavefunctions. In the present 
work, we have used FSRCC theory for one-valence \cite{mani-10,mani-17} to 
compute many-body wavefunctions and the corresponding energies for the ground 
and low-lying excited states of Cs atom. Here, we briefly discuss the theory
used in our calculations. For a one-valence atom or ion, the many-body 
wavefunction is the solution of the eigen value equation
\begin{equation}
  H^{\rm DCB}|\Psi_v \rangle = E_v |\Psi_v \rangle,
  \label{hdc_eqn}
\end{equation}
where $|\Psi_v \rangle$ is the exact wavefunction with the corresponding
energy $E_v$ and $H^{\text{DCB}}$ is the Dirac-Coulomb-Breit no-virtual-pair 
Hamiltonian. For an atom or ion with $N$-electrons
\begin{eqnarray}
	H^{\text{DCB}} & = & \Lambda_{++}\sum_{i=1}^N \left [c\bm{\alpha}_i \cdot
	\mathbf{p}_i + (\beta_i -1)c^2 - V_{N}(r_i) \right ] 
                       \nonumber \\
   & & + \sum_{i<j}\left [ \frac{1}{r_{ij}}  + g^{\text{B}}(r_{ij}) \right] \Lambda_{++},
  \label{ham_dcb}
\end{eqnarray}
where $\bm{\alpha}$ and $\beta$ are the Dirac matrices, and $V_{N}(r_{i})$
is the nuclear potential. The last two terms, $1/r_{ij} $ and
$g^{\text{B}}(r_{ij})$,  are the Coulomb and Breit interactions,
respectively. For Breit interaction, we employ the 
expression \cite{grant-80}
\begin{equation}
  g^{\rm B}(r_{12})= -\frac{1}{2r_{12}} \left [ \bm{\alpha}_1\cdot\bm{\alpha}_2
               + \frac{(\bm{\alpha_1}\cdot \mathbf{r}_{12})
               (\bm{\alpha_2}\cdot\mathbf{r}_{12})}{r_{12}^2}\right].
\end{equation}
The operator $\Lambda_{++}$ projects onto the positive-energy solutions, 
and therefore, sandwiching the Hamiltonian between $\Lambda_{++}$ ensures 
that the effects of the negative-energy continuum states are neglected 
in the present calculations.

\begin{figure}
\begin{center}
  \includegraphics[width = 7.5 cm]{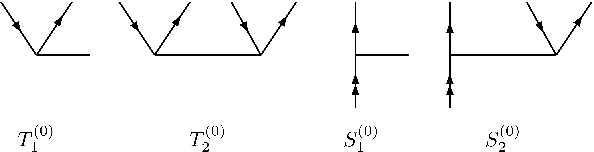}
  \caption{Diagrammatic representation of the unperturbed single and
	  double CC operators for closed-shell and one-valence sectors.}
  \label{ts_fig}
\end{center}
\end{figure}

\begin{figure}
\begin{center}
  \includegraphics[width = 7.0 cm]{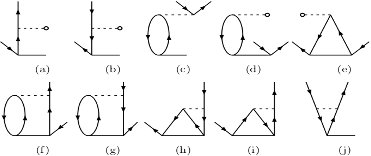}
  \caption{The Goldstone diagrams contributing to the linearized unperturbed
	   CC equation of $T_1^{(0)}$. Diagrams contributing to CC equation of 
	   $S_1^{(0)}$ can be obtained by converting {\em core} line to {\em valence} 
	   line. Dashed lines {\em with} and {\em without}-circle represent 
	   the {\em one-} and {\em two-body} Coulomb interactions, respectively.}
  \label{t10}
\end{center}
\end{figure}

\begin{figure}
\begin{center}
  \includegraphics[width = 7.0 cm]{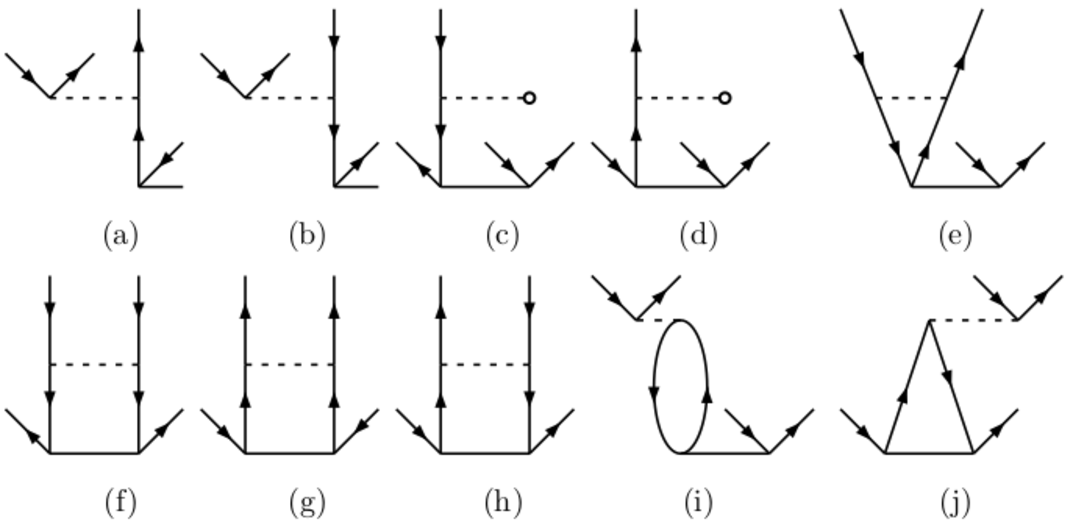}
  \caption{The Goldstone diagrams contributing to the linearized unperturbed
	   CC equation of $T_2^{(0)}$. Diagrams contributing to CC equation of 
	   $S_2^{(0)}$ can be obtained by converting one of {\em core} lines 
	   to a {\em valence} line. Dashed lines {\em with} and {\em without}-circle 
	   represent the {\em one-} and {\em two-body} Coulomb interactions, 
	   respectively.}
  \label{t20}
\end{center}
\end{figure}

In the coupled-cluster (CC) theory, the wavefunction $|\Psi_v \rangle$ is 
expressed in terms of the cluster operators as
\begin{equation}
 |\Psi_v\rangle = e^{T^{(0)}} \left [  1 + S^{(0)} \right ] |\Phi_v\rangle,
 \label{psi_unptrb}
\end{equation}
where $T^{(0)}$ and $S^{(0)}$ represent the unperturbed cluster operators 
for closed-shell and one-valence sectors, respectively. The state
$|\Phi_v\rangle$ is the Dirac-Fock reference state for one-valence system.
It is obtained by adding a valence-electron to the closed-shell reference state
$|\Phi_0\rangle$ using the creation operator $a^\dagger_v$
as $|\Phi_v \rangle = a^\dagger_v|\Phi_0\rangle$. In the coupled-cluster with 
singles and doubles (CCSD) approximation, the $T^{(0)}$ and $S^{(0)}$ 
operators are defined as $T^{(0)} = T^{(0)}_1 + T^{(0)}_2$ and 
$S^{(0)} = S^{(0)}_1 + S^{(0)}_2$, respectively. The expressions of these
operators in the second quantized form are
\begin{subequations}
\begin{eqnarray}
	T^{(0)}_1 &=& \sum_{a, p}t_a^p a_p^{\dagger}a_a, {\;} 
  T^{(0)}_2 = \frac{1}{2!}\sum_{a, b, p, q}t_{ab}^{pq}
  a_p^{\dagger}a_q^{\dagger}a_ba_a, \\
	S^{(0)}_1 &=& \sum_{p}s_v^p a_p^{\dagger}a_v, {\;}
  S^{(0)}_2 = \sum_{a, p, q}s_{va}^{pq}
  a_p^{\dagger}a_q^{\dagger}a_aa_v.
\end{eqnarray}
\end{subequations}
Here, $t_{\cdots}^{\cdots}$ and $s_{\cdots}^{\cdots}$ represent the cluster 
amplitudes and the indices $a$, $b$, $c$, $\ldots$ ($p$, $q$, $r$, $\ldots$) 
represent the core (virtual) states and $v$, $w$, $x$, $\ldots$ represent the 
valence orbitals. The operators $T_1$ ($S_1$) and $T_2$ ($S_2$) generate
single and double electron replacements after operating on the 
closed(open)-shell reference states. The diagrammatic representation 
of these operators are shown in Fig. \ref{ts_fig}. The open-shell 
CC operators $S^{(0)}_1$ and $S^{(0)}_2$  are obtained by solving the 
coupled linear equations \cite{mani-10, mani-17}
\begin{subequations}
\label{s0_eqn}
\begin{eqnarray}
  \langle \Phi_v^p|\bar H_N \! +\! \{\contraction[0.5ex]
  {\bar}{H}{_N}{S} \bar H_N S^{(0)}\} |\Phi_v\rangle
  &=&E_v^{\rm att}\langle\Phi_v^p|S^{(0)}_1|\Phi_v\rangle ,
  \label{s01_eqn}     \\
  \langle \Phi_{va}^{pq}|\bar H_N +\{\contraction[0.5ex]
  {\bar}{H}{_N}{S}\bar H_N S^{(0)}\} |\Phi_v\rangle
  &=& E_v^{\rm att}\langle\Phi_{va}^{pq}|S^{(0)}_2|\Phi_v\rangle.
  \label{s02_eqn}
\end{eqnarray}
\end{subequations}
Here, $H_{\rm N} = H^{\rm DCB} -\langle\Phi_0|H^{\rm DCB}|\Phi_0\rangle$
is the normal ordered Hamiltonian and 
$\bar H_{\rm N}=e^{-T^{(0)}}H_{\rm N}e^{T^{(0)}} $ 
is a similarity transformed Hamiltonian. The energy $E_v^{\rm att}$ is 
the attachment energy of the valence electron, expressed as 
$E_v^{\rm att} = \epsilon_v + \Delta E_v$, where $\epsilon_v$ is the 
single-particle energy and $\Delta E_v$ is the correlation energy.
Similarly, the closed-shell CC operators $T^{(0)}_1$ and $T^{(0)}_2$ are 
the solutions of coupled nonlinear equations \cite{mani-09, mani-17}
\begin{subequations}
\label{t0_eqn}
\begin{eqnarray}
  \langle\Phi^p_a|\bar H_{\rm N}|\Phi_0\rangle = 0,
     \label{t01_eqn}                        \\
  \langle\Phi^{pq}_{ab}|\bar H_{\rm N}|\Phi_0\rangle = 0.
     \label{t02_eqn}
\end{eqnarray}
\end{subequations}
Here, $|\Phi^p_a\rangle$ and $|\Phi^{pq}_{ab}\rangle$ represent the 
Slater determinants with single and double excitations. 
In the Figs. \ref{t10} and \ref{t20}, we have shown the CC diagrams
of the {\em linearized} RCC theory for closed-shell atoms corresponding to
the $T^{(0)}_1$ and $T^{(0)}_2$ equations, respectively. The diagrams 
corresponding to $S^{(0)}_1$ and $S^{(0)}_2$ can be obtained by converting 
one of the {\em core} lines to {\em valence} line.

\subsection{FS-PRCC theory of PNC for one-valence systems}

In the presence of NSD-PNC interaction, the atomic Hamiltonian is modified to
\begin{equation}
    H_{\rm t} = H^{\rm DCB} + \lambda H_{\rm PNC},
    \label{total_H}
\end{equation}
where $\lambda$ is a perturbation parameter and $H_{\rm PNC}^{\rm NSD}$ is the 
NSD-PNC interaction Hamiltonian with the expression
\begin{equation}
   H_{\rm PNC}^{\rm NSD}=\frac{G_{\rm F}\mu'_W}{\sqrt{2}}\sum_i
   \bm{\alpha}_i\cdot \mathbf{I}\rho_{\rm{N}}(r),
  \label{hpnc_nsd}
\end{equation}
where, $G_{\rm F} = 2.22\times10^{-14}$ a.u. is the Fermi coupling constant, 
$\mu'_W$ is the weak nuclear moment of the nucleus, $\bm{\alpha}$
is the Dirac matrix, and ${\mathbf I}$ and $\rho_{\rm{N}}(r)$ are the 
nuclear spin and density, respectively. Defining
$H_{\rm elec}^{\rm NSD} =( G_{\rm F}\mu'_W/\sqrt{2})\sum_i
\bm {\alpha}_i\rho_{\rm{N}}(r)$, the interaction Hamiltonian can be
written in the compact form
\begin{equation}
   H_{\rm PNC}^{\rm NSD}= \helec\cdot \mathbf{I}.
\end{equation}
Due to $H_{\rm PNC}^{\rm NSD}$ the atomic states mix with opposite parity 
states, and hence, these are no longer parity eigen states. We refer to these 
mixed parity states as perturbed states and are solutions of the eigen value 
equation
\begin{equation}
  H_{\rm t} |\widetilde{\Psi}_v \rangle =
  \widetilde{E}_v |\widetilde{\Psi}_v \rangle,
  \label{ht_eqn}
\end{equation}
where $\widetilde{E}_v$ is the perturbed eigen energy. To first-order in 
$\lambda $, $|\widetilde{\Psi}_v \rangle = |\Psi_v\rangle +
\lambda |\bar{\Psi}^{1}_v\rangle$ and $\widetilde{E}_v = E_v 
+ \lambda E^{1}_v$, where the bar in $|\bar{\Psi}^{1}_v\rangle$ denotes a 
state opposite in parity to $|\Psi_v\rangle$.  Using Eqs. (\ref{total_H}) 
and (\ref{hpnc_nsd}) in Eq. (\ref{ht_eqn}), we obtain
\begin{equation}
  \left ( H^{\rm DCB} + \lambda \helec\cdot\mathbf{I} \right) |
  \widetilde{\Psi}_v \rangle = E_v| \widetilde{\Psi}_v \rangle.
  \label{ht_elc_eqn}
\end{equation}
Here, we have used 
$E^1_v = \langle \Psi_v|H_{\rm PNC}^{\rm NSD}|\Psi_v\rangle = 0$, as 
$H_{\rm PNC}^{\rm NSD}$ is an odd parity operator it connects only 
to states with opposite parities.
The presence of nuclear spin operator ${\mathbf I}$ in NSD-PNC 
Hamiltonian leads to two important considerations. First, the cluster 
operators in the electronic sector are rank one operators. And second, 
atomic states are the eigenstates of the total angular 
momentum operator ${\mathbf F} = {\mathbf I} + {\mathbf J}$.

\begin{figure}
\begin{center}
  \includegraphics[width = 7.5 cm]{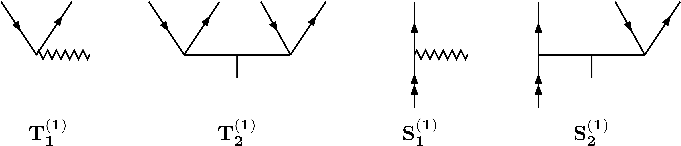}
  \caption{Diagrammatic representations of NSD-perturbed cluster operators
	   in closed-shell and one-valence sectors. The extra line in
	   ${\mathbf T}_2^{(1)} $ and ${\mathbf S}_2^{(1)}$ is to 
	   indicate the multipole structure of the operators.}
  \label{pts_nsd_fig}
\end{center}
\end{figure}

In FS-PRCC theory, the perturbed wavefunction is written as
\begin{equation}
  | \widetilde{\Psi}_v \rangle = e^{T^{(0)}}\left[ 1
    + \lambda \pt \cdot\mathbf{I} \right] \left[ 1
    + S^{(0)} +\lambda \ps \cdot\mathbf{I} \right] |\Phi_v \rangle,
  \label{psi_ptrb}
\end{equation}
where $\pt$ and $\ps$ are  the closed-shell and one-valence perturbed
operators, respectively. Here, the superscript $(1)$ is used to indicate
that the clusters subsume the perturbation. Similar to the case of 
unperturbed operators $T^{(0)}$ and $S^{(0)}$, in the CCSD approximation, we 
consider 
$\mathbf{T}^{(1)} =  {\mathbf T}_1^{(1)} + {\mathbf T}_2^{(1)}$ and
$\mathbf{S}^{(1)} =  {\mathbf S}_1^{(1)} + {\mathbf S}_2^{(1)}$.  
Since these are tensor operators, they satisfy selection rules which are
different from the unperturbed cluster operators. The second quantized 
tensorial representation of ${\mathbf T}^{(1)}$ is described in detail in 
one of our previous works \cite{ravi-20}. Here we discuss the second 
quantized representation and selection rules for the operator 
$\mathbf{S}^{(1)}$. In the second quantized notation
\begin{subequations}
\begin{eqnarray}
   \mathbf{S}_1^{(1)} &= &\sum_{p}\xi_v^p \mathbf {C}_1(\hat r)
            a_p^{\dagger}a_v,
  \label{s1_1}    \\
  \mathbf{S}_2^{(1)}  &= &\sum_{apq} \sum_{lk} \xi_{va}^{pq}(l,k)
  \mathbf{C}_l(\hat r_1)\mathbf{C}_k(\hat r_2) a_p^{\dagger}a_q^
  {\dagger}a_aa_v.
  \label{S2_1}
\end{eqnarray}
\end{subequations}
Here, $\xi^{\cdots}_{\cdots}$ represents the cluster amplitude for the operator
$\mathbf{S}^{(1)}$. The one-body operator ${\mathbf S}_1^{(1)}$ is an odd
parity operator and expressed in terms of a rank-one $\mathbf{C}$-tensor. It
satisfies the parity and triangular selection rules $(-1)^{l_v+l_p} = -1$ and  
$|j_v-j_p| \leqslant 1 \leqslant (j_v + j_p)$. The tensor structure of the 
two-body operator $\mathbf{S}_2^{(1)}$ involves two $\mathbf{C}$-tensors of 
ranks $l$ and $k$ associated with its two-vertices. These two 
$\mathbf{C}$-tensors couple to give a rank-one operator, $\mathbf{S}_2^{(1)}$. 
The selection rules of $\mathbf{S}_2^{(1)}$ are
$(-1)^{l_v+l_p} = -(-1)^{l_a+l_q}$ and
$|j_v-j_p| \leqslant l \leqslant (j_v+j_p)$,
$|j_a-j_q| \leqslant k \leqslant (j_a+j_q)$. The diagrammatic
representations of $\mathbf{T}^{(1)}$ and $\mathbf{S}^{(1)}$ are 
shown in Fig. \ref{pts_nsd_fig}.

\begin{figure}
\begin{center}
  \includegraphics[width = 7.0 cm]{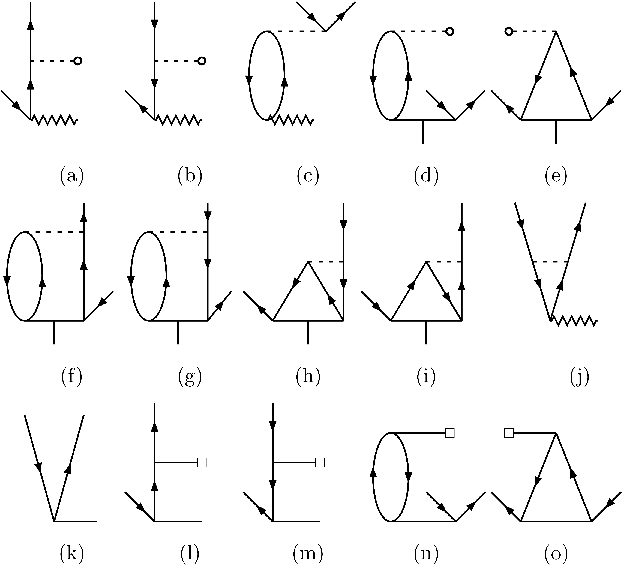}
  \caption{The Goldstone diagrams which contribute to the linearized perturbed
	   CC equations of ${\mathbf T}_1^{(1)}$. The diagrams for 
           ${\mathbf S}_1^{(1)}$ can be obtained by converting {\em core} line 
           to the {\em valence} line. The dashed lines {\em with} and 
           {\em without}-circle represent the {\em one-} and {\em two-body} 
           Coulomb interactions, respectively. The bar with square represents 
           the PNC operator.}
  \label{t11}
\end{center}
\end{figure}

\begin{figure}
\begin{center}
  \includegraphics[width = 7.0 cm]{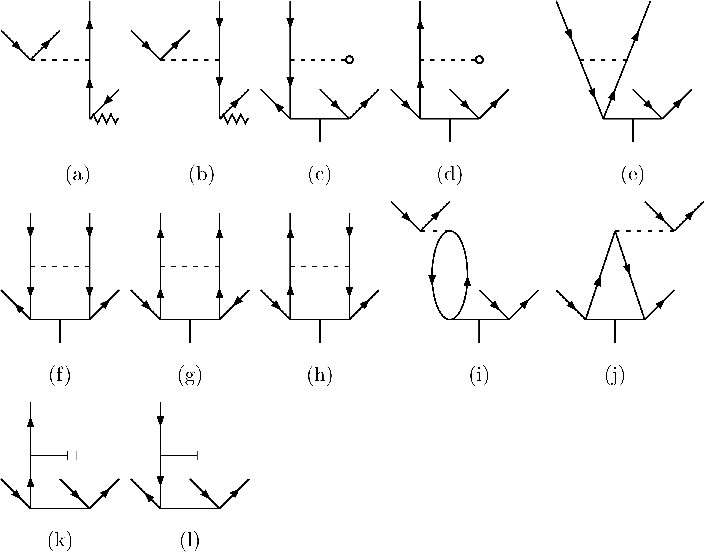}
  \caption{The Goldstone diagrams which contribute to the linearized perturbed
	   CC equations of ${\mathbf T}_2^{(1)}$. The diagrams for 
           ${\mathbf S}_2^{(1)}$ can be obtained by converting {\em core} line 
           to the {\em valence} line. The dashed lines {\em with} and 
           {\em without}-circle represent the {\em one-} and {\em two-body} 
           Coulomb interactions, respectively. The bar with square represents 
           the PNC operator.}
  \label{t21}
\end{center}
\end{figure}

Using Eq. (\ref{psi_ptrb}) in Eq. (\ref{ht_elc_eqn}) and projecting entire 
equation on $e^{-T^{(0)}}$ from left and retaining the terms linear 
in $\lambda$, we get 
\begin{eqnarray}
  \Big[ \bar H_{\rm N}\ps  &+& \bar H_{\rm N}\pt ( 1 + S^{(0)} ) +
    \bar {\mathbf{H}}_{\rm elec}^{\rm NSD} ( 1 + S^{(0)} ) \Big]
    |\Phi_v \rangle      \nonumber \\
  &=&\left[ \Delta E_v \ps + \Delta E_v \pt ( 1 + S^{(0)} ) \right]|\Phi_v 
     \rangle. 
  \label{deltae1v}
\end{eqnarray}
Here, $\Delta E_v = E_v - \langle\Phi_v|H^{\rm DCB}|\Phi_v\rangle$ is
the correlation energy of the one-valence system. And, like $\bar{H}_{\rm N}$ 
introduced earlier, $\bar {{\mathbf H}}_{\rm elec}^{\rm NSD} =
e^{-T^{(0)}}\helec e^{T^{(0)}}$ is the similarity transformed PNC Hamiltonian 
in the electronic space. The PRCC equations of the ${\mathbf S}^{(1)}_1$ and 
${\mathbf S}^{(1)}_2$ can be obtained by projecting Eq. (\ref{deltae1v}) with 
single and double excited determinants $\langle\Phi^p_v|$ and 
$\langle\Phi^{pq}_{va}|$, respectively, as
\begin{widetext}
\begin{subequations}
\begin{eqnarray}
   \langle \Phi^p_v |\{ \contraction[0.5ex]{}{H}{_{\rm N}}{S}\bar{H}_{\rm N}
	\mathbf{S}^{(1)} \} + \{ \contraction[0.5ex]{}{H}{_{\rm N}}{S}\bar{H}_{\rm N}
     \mathbf{T}^{(1)} \} + \{ \contraction[0.5ex]{}{H}{_{\rm N}}{T}
     \contraction[0.8ex]{}{V}{_{\rm N}T^{(1)}}{S}\bar{H}_{\rm N}
     \mathbf{T}^{(1)}S^{(0)}\} + \bar{\mathbf{H}}_{\rm elec}^{\rm NSD}
    + \{ \contraction[0.5ex]{}{H}{_{\rm elec}^{\rm NSD}}{S}
     \bar{\mathbf{H}}_{\rm elec}^{\rm NSD}{S}^{(0)} \}|\Phi_v \rangle &=&
  E_v^{\rm att} \langle \Phi^p_v | \pso|\Phi_v \rangle, \\
  \langle \Phi^{pq}_{va}|\{ \contraction[0.5ex]{}{H}{_{\rm N}}{S}\bar{H}_{\rm N}
     \mathbf{S}^{(1)}\}+\{ \contraction[0.5ex]{}{H}{_{\rm N}}{S}\bar{H}_{\rm N}
     \mathbf{T}^{(1)} \} + \{ \contraction[0.5ex]{}{H}{_{\rm N}}{T}
     \contraction[0.8ex]{}{V}{_{\rm N}T^{(1)}}{S}\bar{H}_{\rm N}
     \mathbf{T}^{(1)}S^{(0)}\} + \bar{\mathbf{H}}_{\rm elec}^{\rm NSD}
    + \{ \contraction[0.5ex]{}{H}{_{\rm elec}^{\rm NSD}}{S}
     \bar{\mathbf{H}}_{\rm elec}^{\rm NSD}{S}^{(0)} \}|\Phi_v \rangle &=&
   E_v^{\rm att} \langle \Phi^{pq}_{va} | \pst|\Phi_v \rangle.
  \label{ccsptrb1v2}
\end{eqnarray}
\label{prcc_eqn}
\end{subequations}
\end{widetext}
In deriving the above equations we have used the relations,
$ \langle \Phi^p_v | \pt |\Phi_v \rangle = 0$ and
$\langle \Phi^p_v| \pt S^{(0)}| \Phi_v \rangle = 0$. These follow because
$\pt$, as an operator of closed-shell sector, does not contribute to the
PRCC equation of $\pso$ and $\pst$. The closed-shell perturbed operators 
$\pt$ are the solutions of a similar set of coupled 
equations \cite{mani-09, ravi-20}
\begin{subequations}
\label{pcceq}
\begin{eqnarray}
  \langle \Phi^p_a |\{ \contraction{}{H}{_{\rm N}}{T}
     \bar{\mathbf{H}}_{\rm N}\mathbf{T}^{(1)} \} |\Phi_0\rangle &=&
  -\langle \Phi^p_a | \bar {\mathbf{H}}_{\rm elec}^{\rm NSD}
      - \Delta E_0 \pt |\Phi_0 \rangle, \;\;\;\;\;\;\;\;
  \label{pcceq1}                         \\
  \langle \Phi^{pq}_{ab} | \{\contraction{}{H}{_{\rm N}}{T}
  \bar{\mathbf{H}}_{\rm N}\mathbf{T}^{(1)} \} |\Phi_0 \rangle &=&
  -\langle \Phi^{pq}_{ab} | \bar {\mathbf{H}}_{\rm elec}^{\rm NSD}
   - \Delta E_0 \pt  |\Phi_0 \rangle,
 \label{pcceq2}
\end{eqnarray}
\end{subequations}
where $\Delta E_0$ is the correlation energy of closed-shell sector of the 
atom. In Figs. \ref{t11} and \ref{t21}, we have shown the CC diagrams for 
$\mathbf{T}^{(1)}_1$ and $\mathbf{T}^{(1)}_2$ which contribute to 
the {\em linearized} PRCC theory. The corresponding diagrams
for $\mathbf{S}^{(1)}_1$ and $\mathbf{S}^{(1)}_2$ are obtained by transforming
one of the {\em core} lines to a {\em valence} line. 

\subsection{Calculation of $E1^{\rm NSD}_{\rm PNC}$ using FS-PRCC theory}

The PNC induced electric dipole transition amplitude $E1^{\rm NSD}_{\rm PNC}$
is the transition matrix element of the dipole operator $\mathbf{D}$ between 
the initial and final perturbed states
\begin{equation}
   E1^{\rm NSD}_{\rm PNC} = \frac{\langle \widetilde{\Psi}_w\red \mathbf{D}\red
     \widetilde{\Psi}_v \rangle} {\sqrt{\langle \Psi_v |\Psi_v \rangle} 
     \sqrt{\langle \Psi_w| \Psi_w \rangle}}.
\label{e1nsd}
\end{equation}
The advantage with this expression is that, unlike the often used 
sum-over-states approach \cite{gopakumar-07}, it implicitly accounts for all 
possible intermediate states. Using Eq. (\ref{psi_ptrb}) in (\ref{e1nsd}) and 
retaining terms up to second order in CC operators, we can write the 
electronic component of $E1^{\rm NSD}_{\rm PNC}$, with the CCSD approximation, 
as
\begin{widetext}
\begin{eqnarray}
E1_{\rm elec}^{\rm NSD} & \approx & \frac{1}{\cal N} \langle \Phi_w |
            \left ( {\mathbf D}\mathbf{S}_1^{(1)} +{\mathbf D}
            \mathbf{S}_2^{(1)} + S_1^{(0)\dagger}{\mathbf D} \mathbf{S}_1^{(1)}
            + S_1^{(0)\dagger}{\mathbf D}\mathbf{S}_2^{(1)}
            + S_2^{(0)\dagger}{\mathbf D} \mathbf{S}_1^{(1)}
            + S_2^{(0)\dagger}{\mathbf D} \mathbf{S}_2^{(1)}
            +  S_1^{(0)\dagger}{\mathbf D} \mathbf{T}_1^{(1)}
            + S_2^{(0)\dagger}{\mathbf D} \mathbf{T}_1^{(1)} \right.
                     \nonumber  \\
       &&   \left. + S_2^{(0)\dagger}{\mathbf D} \mathbf{T}_2^{(1)}
            + T_1^{(0)\dagger}{\mathbf D}\mathbf{S}_1^{(1)}
            + T_1^{(0)\dagger}{\mathbf D} \mathbf{S}_2^{(1)}
            + T_2^{(0)\dagger} {\mathbf D}\mathbf{S}_2^{(1)}
            + {\mathbf D} \mathbf{T}^{(1)}
            + T_1^{(0)\dagger} {\mathbf D}\mathbf{T}_1^{(1)}
            + T_1^{(0)\dagger} {\mathbf D} \mathbf{T}_2^{(1)}
            + T_2^{(0)\dagger} {\mathbf D} \mathbf{T}_1^{(1)} \right.
                      \nonumber  \\
       && \left. + T_2^{(0)\dagger} {\mathbf D} \mathbf{T}_2^{(1)} \right)
            + {\text{H.c.}} + {\mathbf D} \mathbf{T}_1^{(1)} S_1^{(0)}
            + {\mathbf D} \mathbf{T}_1^{(0)}\mathbf{S}_1^{(1)} |\Phi_v \rangle.
  \label{e1nsd2}
\end{eqnarray}
\end{widetext}
Here,
\begin{eqnarray}
   {\cal N} = \sqrt{\langle \Phi_v|(e^{T^{(0)}}(1 + S^{(0)}))^\dagger
   (e^{T^{(0)}}(1 + S^{(0)}))| \Phi_v \rangle} \nonumber \\
	\times \sqrt{\langle \Phi_w|(e^{T^{(0)}}(1 + S^{(0)}))^\dagger
   (e^{T^{(0)}}(1 + S^{(0)}))| \Phi_w\rangle},
\end{eqnarray}
is the normalization factor.
The terms $S_1^{(0)\dagger} {\mathbf D} \mathbf{T}_2^{(1)} + \text{H.c.}$,
$T_2^{(0)\dagger} {\mathbf D} \mathbf{S}_2^{(1)} + \text{H.c.}$ and
${\mathbf D}\mathbf{T}_2^{(1)} + \text{H.c.}$ are not included as 
these do not contribute to $E1_{\rm elec}^{\rm NSD}$ for one-valence system.

\begin{figure}[ht]
\begin{center}
  \includegraphics[width = 7.0cm]{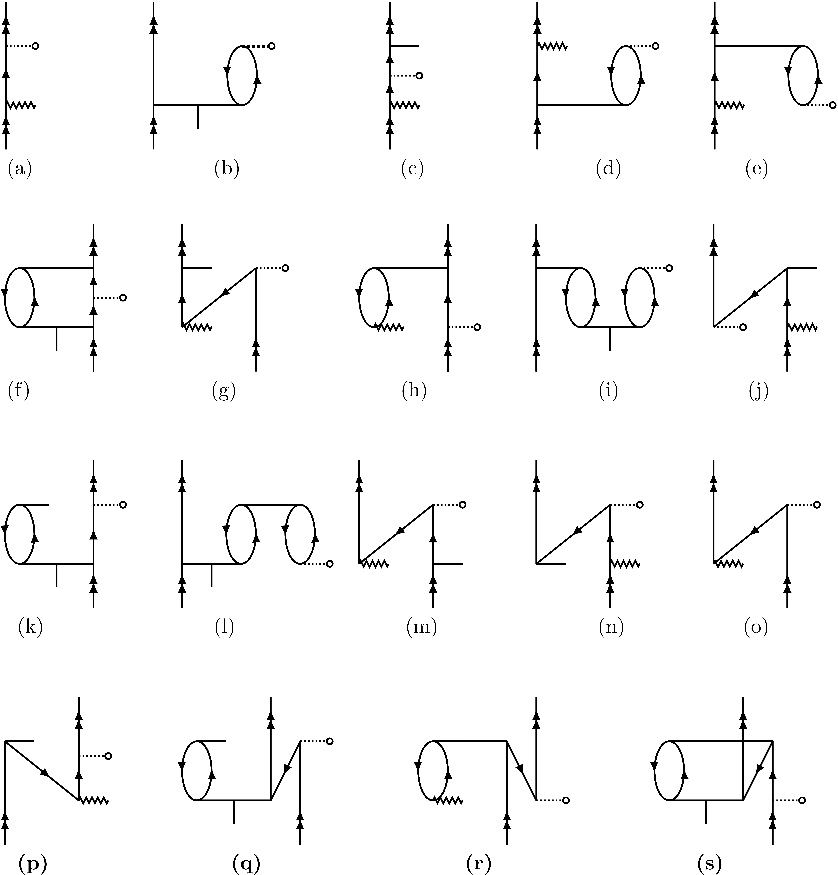}
  \caption{Some example Goldstone diagrams contributing to 
	$E1_{\rm elec}^{\rm PNC}$ for one-valence systems 
	in Eq. (\ref{e1nsd2}).}
  \label{e1pnc_cc_fig}
\end{center}
\end{figure}

As evident, the expression in Eq. (\ref{e1nsd2}) is independent of the 
nuclear spin, and hence can be computed within the electronic space alone. 
To evaluate this, we use Goldstone diagrams to identify the 
contributing terms, and these are then evaluated using the diagrammatic
techniques \cite{lindgren-86}. As demonstrated in our previous 
work \cite{mani-10}, the diagrams arising from terms with cubic or higher 
orders of CC operators have negligible contributions, so they are not 
included in the equation. There are 128 Goldstone diagrams which contribute 
to Eq. (\ref{e1nsd2}) and, as an example, we show one diagram from each of 
the terms in Eq. (\ref{e1nsd2}) in Fig. \ref{e1pnc_cc_fig}. The diagrams 
from the Hermitian conjugate terms are not shown as these are topologically 
equivalent. Among all the terms, the first two terms 
${\mathbf D} \mathbf{S}_1^{(1)}$ and ${\mathbf D} \mathbf{S}_2^{(1)}$, and 
their Hermitian conjugates, are expected to have the dominant contribution 
to $E1_{\rm elec}^{\rm NSD}$. The reason for this can be attributed to the 
larger magnitudes of the one-valence CC operators and the strong effect of 
perturbation on these operators.

\subsubsection{Coupling with nuclear spin}

As discussed earlier, the PNC diagrams, shown in Fig. \ref{e1pnc_cc_fig}, are 
defined in the electronic subspace. However, to calculate 
$E1^{\rm NSD}_{\rm PNC}$ the nuclear spin part of the PNC operator must be 
coupled with the operators in the electronic sector. For this reason
the $E1^{\rm NSD}_{\rm PNC}$ diagrams must include a vertex which operates on 
the nuclear spin states. This coupling with nuclear part is done using the 
angular momentum algebra by following the conventions described 
in Lindgren and Morrison \cite{lindgren-86}.  

\begin{figure}
\begin{center}
  \includegraphics[width = 7.0cm]{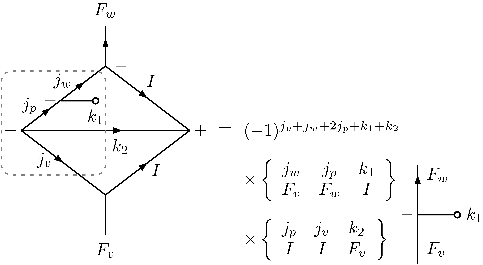}
	\caption{Diagram showing the coupling of nuclear spin with electronic 
	state and resulting angular factor. Portion within the rectangled dash 
	line represents the electronic part, whereas the remaining portion is 
        due to the coupling of nuclear spin and electronic angular momenta.}
  \label{ei_1}
\end{center}
\end{figure}

To elaborate on this, we choose one of the simple diagrams shown in 
Fig. \ref{e1pnc_cc_fig}(a) as an example. It arises from the CC operator
${\mathbf S}^{(1)}_1$, and subsumes the Dirac-Fock and dominant 
core-polarization contributions. Together these contributions make it the most 
dominant diagram contributing to $E1_{\rm elec}^{\rm NSD}$. The angular 
momentum diagram which represents coupling with the nuclear spin is shown in 
Fig. \ref{ei_1}. Here, $j_v$, $j_w$ and $j_p$ represent the total angular 
momenta of the single particle states. The line $k_1$ represents a rank one 
multipole operator denoting the angular part of dipole operator. Similarly,
$k_2$ is another rank one multipole operator which connects the electronic 
and nuclear sectors. Since ${\mathbf I}$ is a diagonal operator, $k_2$ does 
not change the nuclear spin, whereas in the electronic sector it facilitates 
a transition from $j_v$ to $j_p$ states. The electronic and nuclear angular 
momenta ($j_v$, ${\mathbf I}$) and ($j_w$, ${\mathbf I}$) are coupled to give 
hyperfine states $|F_v\rangle$ and $|F_w\rangle$, respectively. Using the 
Wigner-Eckart theorem, we can write the algebraic expression for the coupled 
diagram, shown in Fig. \ref{ei_1}, as
\begin{eqnarray}
	&&\langle F_w m_w|{\mathbf D} {\mathbf S}^{(1)}_1\cdot {\mathbf I}
         | F_v m_v\rangle = (-1)^{F_w - m_w} \nonumber \\
	&&\times 
  \left(\begin{array}{ccc}
	  F_w & 1 & F_v\\
	  -m_w & q & m_v\\
       \end{array}\right) 
	\langle F_w||{\mathbf D}_{\rm eff}||F_v \rangle.
\end{eqnarray}
Here, ${\mathbf D}_{\rm eff} = {\mathbf D}({\mathbf S}^{(1)}_1\cdot 
{\mathbf I})$ is a rank one operator, $m_i$ represents the hyperfine magnetic 
quantum number and $q$ is the component of ${\bf D}_{\rm eff}$. The phase 
factor and $3j$-symbol correspond to the diagram on the right hand side 
of equation in Fig. \ref{ei_1}. 

\section{Results and discussions}
\label{results}
 
\subsection{Single-particle basis and convergence of properties}

Accurate single-electron basis is essential to get reliable results using 
FSRCC theory. In this work, we have used Gaussian-type orbitals 
(GTOs) \cite{mohanty-91} as the single-electron basis in all the 
calculations. The GTOs are expressed as a linear combination of 
the Gaussian-type functions (GTFs) and the GTFs used in defining the large 
component of the wavefunction are 
\begin{equation}
  g^L_{\kappa p} (r) = C^L_{\kappa i} r^{n_\kappa} e^{-\alpha_p r^2}.
\end{equation}
Here, parameter $p = 0$, $1$, $2$, $\ldots$, $N$ is the GTO index and 
$N$ is the total number of GTFs. The exponent $\alpha_p$ is further 
expressed in terms of two independent parameters $\alpha_{0}$ and $\beta$, 
as $\alpha_0 \beta^{p-1}$. Parameters $\alpha_{0}$ and $\beta$ are optimized 
separately for each orbital symmetry so that the single-electron wavefunctions 
and the corresponding energies match well with the numerical values obtained 
from GRASP2K \cite{jonsson-13}. To ensure a good quality basis, we have also
compared our gaussian energies with B-spline energies \cite{oleg-16}. 
The small components of wavefunctions are derived from the large components 
using the kinetic balance condition \cite{stanton-84}.

\begin{table}
\begin{center}
\caption{Orbital energies (in hartree) from GTO using Dirac-Coulomb Hamiltonian 
	compared with GRASP2K \cite{jonsson-13} and B-Spline \cite{oleg-16} results. 
	The optimized $\alpha_0$ and $\beta$ parameters for even tempered GTO basis 
	used in the calculations are also provided.}
\begin{ruledtabular}
\begin{tabular}{cccc}
  Orbitals  &  GRASP2K   & B-spline  & GTO \\
\hline
  $1s_{1/2} $ & $1330.11688 $ & $1330.11804 $ & $1330.11562$\\
  $2s_{1/2} $ & $212.56408  $ & $212.56447  $ & $212.56407 $\\
  $2p_{1/2} $ & $199.42926  $ & $199.42946  $ & $199.42939 $\\
  $2p_{3/2} $ & $186.43638  $ & $186.43658  $ & $186.43658 $\\
  $3s_{1/2} $ & $45.96965   $ & $45.96973   $ & $45.96966  $\\
  $3p_{1/2} $ & $40.44825   $ & $40.44829   $ & $40.44828  $\\
  $3p_{3/2} $ & $37.89430   $ & $37.89430   $ & $37.89431  $\\
  $3d_{3/2} $ & $28.30949   $ & $28.30949   $ & $28.30950  $\\
  $3d_{5/2} $ & $27.77517   $ & $27.77515   $ & $27.77516  $\\
  $4s_{1/2} $ & $9.51283    $ & $9.51282    $ & $9.51280   $\\
  $4p_{1/2} $ & $7.44630    $ & $7.44628    $ & $7.44628   $\\
  $4p_{3/2} $ & $6.92101    $ & $6.92100    $ & $6.92100   $\\
  $4d_{3/2} $ & $3.48563    $ & $3.48561    $ & $3.48562   $\\
  $4d_{5/2} $ & $3.39691    $ & $3.39690    $ & $3.39690   $\\
  $5s_{1/2} $ & $1.48981    $ & $1.48980    $ & $1.48980   $\\
  $5p_{1/2} $ & $0.90789    $ & $0.90789    $ & $0.90789   $\\
  $5p_{3/2} $ & $0.84033    $ & $0.84033    $ & $0.84033   $\\
\hline
$E_{\rm SCF}$ & $7786.64263 $ & $7786.64638 $ & $7786.63884$ \\
\hline
  Parameter   &  $s$   & $p$  &  $d$ \\
\hline
  $\alpha_0$ & 0.00340 & 0.00368 & 0.00495\\
  $\beta$  & 1.779  & 1.766  &  1.829 \\
 \end{tabular}
\end{ruledtabular}
\label{tab_eps}
\end{center}
\end{table}

In Table \ref{tab_eps}, we have provided the optimized single-electron and 
the self-consistent-field (SCF) energies for Cs and have compared them with
GRASP2K and B-spline results. We have also tabulated the optimized values 
of the $\alpha_0$ and $\beta$ parameters. It is to be noted that, the 
single-electron basis used in the properties calculations incorporates 
the effects of vacuum polarization and the self-energy corrections. As 
evident from the table, the single-particle and SCF energies are in excellent 
agreement with the GRASP2K and B-spline results. The difference observed is 
of the order of {\em milli-hartree} or less.

\begin{figure}
\begin{center}
  \includegraphics[scale = 0.35, angle=-90]{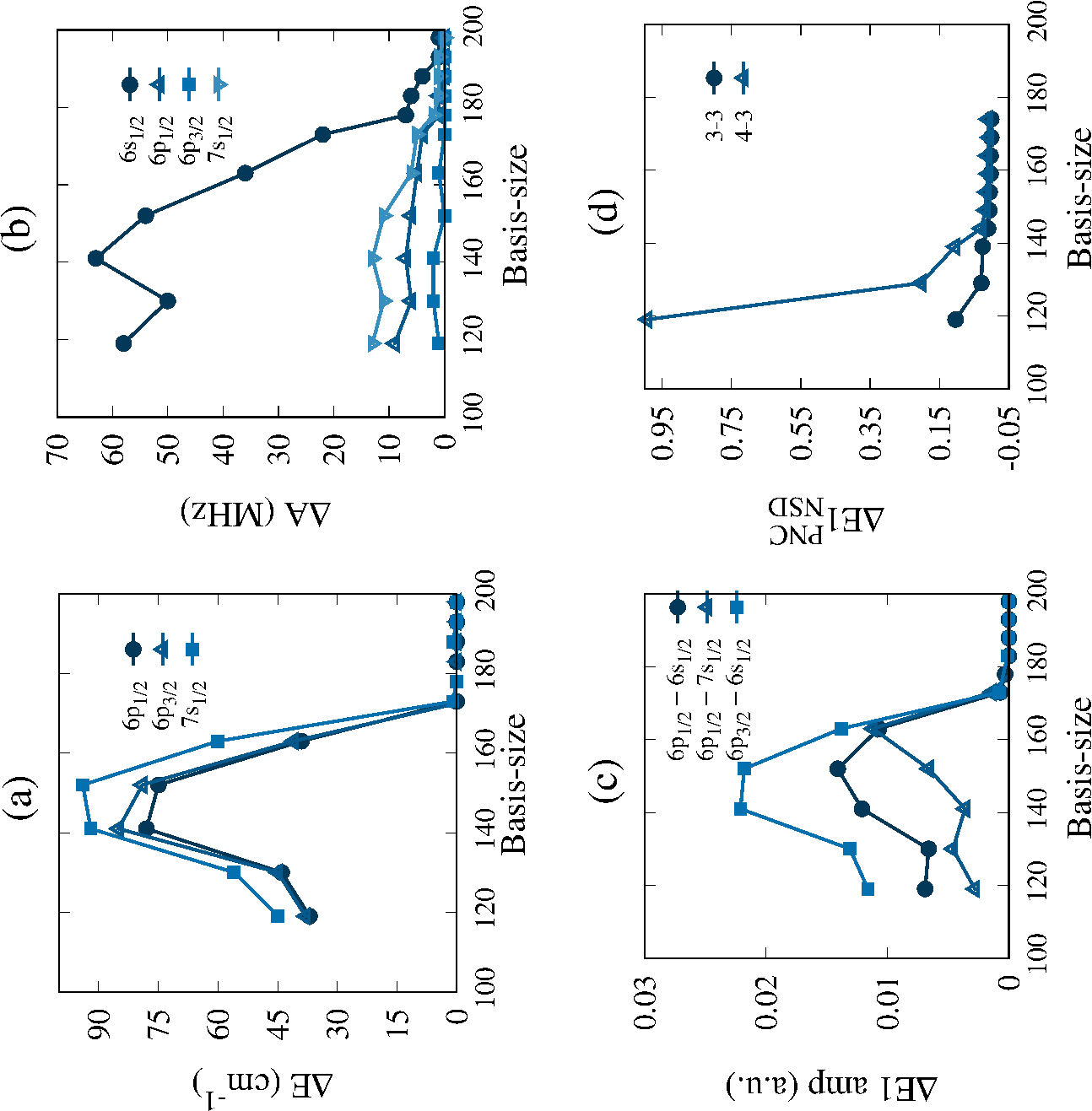}
  \caption{Convergence of the properties results computed using FS-RCC 
	theory with basis size.}
  \label{fig_conv}
\end{center}
\end{figure}

Owing to the mathematically incomplete nature of GTOs, a check on the 
convergence of properties with basis size is essential to obtain reliable 
results using FSRCC. To assess the convergence of properties, in Tables 
\ref{basis_conv} and \ref{basis_e1pnc} of Appendix, we have listed the 
values of excitation energies, $E1$ matrix elements, HFS constants, and 
$E1^{\rm NSD}_{\rm PNC}$ as a function of increasing basis sizes. As 
discernible from the table, to obtain a converged basis set, we start with a 
moderate-size basis and add orbitals systematically to each symmetry until 
the change in the properties is less than or equal to $10^{-3}$ in the
respective units. To visualize the trend, in Fig. \ref{fig_conv}, we show 
the convergence of excitation energies, HFS constants, and $E1$ matrix elements
with the basis size. As discernible from the 
panel (a) of Fig. \ref{fig_conv}, all the properties converge well with 
the basis size. For an example, the change in the excitation 
energies is almost zero when the basis is augmented beyond 163 
($19s$, $17p$, $16d$, $15f$, $13g$, $11h$) orbitals. So, we consider this as 
the optimized basis set and use it in the calculations by adding the 
corrections from relativistic and QED effects.

\subsection{Excitation energies, hyperfine structure constants and E1 
            transition amplitudes}

\begin{table}
	\caption{Excitation energies (cm$^{-1}$) of some low lying states of Cs.}
        \label{orbe_cs}
        \begin{ruledtabular}
           \begin{tabular}{cccccccc}
                \multicolumn{1}{c}{\text{States}}          &
                \multicolumn{1}{c}{\text{DC-CCSD}}     &
                \multicolumn{1}{c}{\text{Breit}}           &
                \multicolumn{1}{c}{\text{QED}}    &

                \multicolumn{1}{c}{\text{Total}}     &
                \multicolumn{1}{c}{\text{NIST \cite{nist-asd}}} &
                \multicolumn{1}{c}{\text{\% diff.}}  \\
                \hline
                $6p_{1/2}$ & 11168 & 10  & 8   & 11186  & 11178 & 0.07 \\
                $6p_{3/2}$ & 11746 & 11  & 8   & 11765  & 11732 & 0.28 \\
                $7s_{1/2}$ & 18503 & 12  & 10  & 18535  & 18535 & 0.00 \\
                $7p_{1/2}$ & 21732 & 13  & 10  & 21755  & 21765 & 0.04 \\
                $7p_{3/2}$ & 21920 & 14  & 12  &  21946 & 21946 & 0.00 \\
           \end{tabular}
        \end{ruledtabular}
	\label{tab_ee}
\end{table}

In this section we present and discuss the excitation energies, MHFS constants, 
and the $E1$ transition amplitudes obtained from our computations. The 
excitation energies of a few low lying states of Cs from our computations 
are listed in Table \ref{tab_ee}. To improve the accuracy of the excited
energies, we have incorporated Breit interaction, self-energy and vacuum 
polarization corrections. The combined corrections from self-energy and vacuum 
polarization are listed as QED contribution. It is to be noted that, obtaining
accurate energies indicates accuracy of the wavefunctions, which are used 
later to compute various properties. As evident from the table, the computed 
excitation energies are in excellent agreement with the NIST 
data \cite{nist-asd}. Among the excited states, the $6p_{3/2}$ state has
the largest deviation, our result is $\approx 0.3$\% larger than the NIST
data. It is, however, to be noted that the deviations of the excited states 
$6p_{1/2}$, $7s_{1/2}$, and $7p_{1/2}$, which have dominant contributions to 
the PNC matrix elements, are less than 0.1\%. To be specific, our result of 
$6p_{1/2}$ excitation energy is $\approx 0.07$\% larger, whereas excitation 
energy of $7s_{1/2}$ is identical with the NIST data. Considering the 
contributions from the Breit and QED corrections, the contributions 
are at the maximum $\approx 0.09$ and $0.07$\%, respectively, which
occurs in the case of $6p_{1/2}$ state.

\begin{table*}[!ht]
   \begin{center}
   \caption{Magnetic dipole hyperfine structure constants (MHz) for some low 
            lying states and E1 transition amplitudes (a.u.) for allowed 
            transitions in Cesium. The data from experiments and other theory 
            calculations are also provided for comparison.}
   \footnotesize
   \begin{ruledtabular}
           \begin{tabular}{lcccccc}
                   \multicolumn{1}{c}{\textrm {States/Transition}}    &
                   \multicolumn{1}{c}{\text{DC-CCSD}} &
                   \multicolumn{1}{c}{\text{Breit }}  &
                   \multicolumn{1}{c}{\text{QED }}  &
                   \multicolumn{1}{c}{\text{Total }}  &
                   \multicolumn{1}{c}{\text{Other cal. }}  &
                   \multicolumn{1}{c}{\text{Exp.}} \\
		   \hline
		   \multicolumn{1}{c}{}&\multicolumn{1}{c}{}&\multicolumn{4}{c}{\text{MHFS ($A$)}} \\
    \hline
   $6s_{1/2}$ &  2297.96 & 6.08 & 4.82 & 2308.86
		   &  $2280.6^{\rm a}$, $2306(10)^{\rm b}$, $2306.6^{\rm c}$, $2302^{\rm d}$
		   &  $2298.16^{\rm h}$, $2298.2^{\rm m}$  \\
		   & & & & & $2300.3^{\rm e}$, $2291^{\rm f}$, $2293.3^{\rm g}$\\

   $6p_{1/2}$ & 289.77   & -2.00 & 0.05  & 287.82
                   & $291 ^{\rm b}$, $291.49^{\rm c}$, $293.5^{\rm d}$
		   & $291.929(1)^{\rm l}$, $291.91^{\rm o}$, $295.02^{\rm i}$ \\
		   & & & & & $290.5^{\rm e}$, $292.7^{\rm f}$, $294.96^{\rm g}$\\
   $6p_{3/2}$ &  51.01   & -1.24 & 0.02 & 49.79
                   & $51.2^{\rm d}$, $49.8^{\rm f}$
		   & $47.19^{\rm i}$, $50.29^{\rm k}$ \\
   $7s_{1/2}$ &  544.44  & -0.31 & 1.08  & 545.21
                   & $547 ^{\rm b}$, $544.59^{\rm c}$, $546.8^{\rm d}$
		   & $545.87(1)^{\rm p}$, $545.82^{\rm q}$, $545.9^{\rm r}$   \\
		   & & & & & $543.8^{\rm e}$, $544.0^{\rm f}$, $545.67(40)^{\rm g}$ & $568.42^{\rm i}$, $546.3^{\rm h}$ \\
   $7p_{1/2}$ & 93.61  & -0.59 & 0.01 & 93.03
                   &  $94(1)^{\rm b}$, $94.07^{\rm c}$, $94.0^{\rm d}$,
		   & $94.40(5)^{\rm j}$,$94.35(4)^{\rm s}$  \\
		   & & & & &  $94.1^{\rm e}$, $94.21^{\rm f} $, $94.49(26)^{\rm g}$ \\
   $7p_{3/2}$ & 16.81  & -0.38 & 0.01 & 16.44
                   & $17.1^{\rm d}$, $16.255^{\rm f}$
		   & $16.605(6)^{\rm j}$ \\
      \hline
      \multicolumn{1}{c}{}  & \multicolumn{1}{c}{}  & \multicolumn{4}{c}{\text{E1 amplitudes}} \\
      \hline
      $6p_{1/2}\rightarrow 6s_{1/2}$   &  4.5378 & 0.0 & -0.0006 & 4.5372
              & $4.5487^{\rm y}$, $4.5093^{\rm c}$, $4.5052^{\rm z}$, $4.528^{\rm a}$
              & $4.5010^{\rm \nu}$, $4.508^{\rm \kappa}$, $4.510^{\rm \lambda}$ \\
      $7p_{1/2}\rightarrow 6s_{1/2}$   & 0.2902 & 0.0 & 0.0003  & 0.2905
                                       & $0.2769^{\rm c}$, $0.3006^{\rm y}$, $0.2776^{\rm z}$
		                       & $0.2781(45)^{\rm \theta}$      \\
      $6p_{3/2}\rightarrow 6s_{1/2}$   & 6.4036 & 0.0 & -0.0007 & 6.4029
             &  $6.3402^{\rm z}$
	     &  $6.345^{\rm \kappa}$, $6.347^{\rm \lambda}$, $6.3349(48)^{\rm \nu}$ \\
      $7p_{3/2}\rightarrow 6s_{1/2}$   & 0.6075 & -0.0001 & -0.0001 & 0.6073
                   & $0.5741^{\rm z}$
		   & $0.5742(57)^{\rm \theta}$     \\
      $6p_{1/2}\rightarrow 7s_{1/2}$   & 4.2601  & 0.0 & 0.0007 & 4.2608
                   & $4.2500^{\rm y}$, $4.239^{\rm z}$, $4.245^{\rm c}$, $4.243^{\epsilon}$
		   & $4.249(4)^{\rm \zeta}$    \\
      $7p_{1/2}\rightarrow 7s_{1/2}$   & 10.3165 & -0.0001 & -0.0011 & 10.3153
                   & $10.297^{\rm y}$, $10.297^{\rm z}$, $10.307^{\rm c}$ 
		   & $10.308(15)^{\rm \zeta}$      \\
      $6p_{3/2}\rightarrow 7s_{1/2}$   & 6.5189 &  0.0 & 0.0009 & 6.5198
                   & $6.507^{\rm \beta}$, $6.474^{\rm z}$, $6.480^{\rm \epsilon}$
                   & $6.489^{\rm \zeta}$      \\
      $7p_{3/2}\rightarrow 7s_{1/2}$   & 14.3267 & -0.0001 & -0.0017 & 14.3249
                   & $14.303^{\rm z}$
                   &    \\
   \end{tabular}
   \end{ruledtabular}
   \begin{flushleft}
           $^{\rm a}$ Ref. {\cite{derevianko-05}} - CC,
           $^{\rm b}$ Ref. {\cite{sahoo-21}} -  RCC,
           $^{\rm c}$ Ref. {\cite{porsev-10}} - CCSDvT,
           $^{\rm d}$ Ref. {\cite{kozlov-01b}} - MBPT,
           $^{\rm e}$ Ref. {\cite{dzuba-02}} - All-order,
           $^{\rm f}$ Ref. {\cite{blundell-91}} - LRCC,
           $^{\rm g}$ Ref. {\cite{grunefeld-19}} - All-order CP,
           $^{\rm h}$ Ref. {\cite{gupta-73}} - Exp.,
           $^{\rm i}$ Ref. {\cite{belin-76}} - Exp.,
           $^{\rm j}$ Ref. {\cite{williams-18}} - Exp.,
           $^{\rm k}$ Ref. {\cite{gerginov-03}} - Exp.,
           $^{\rm l}$ Ref. {\cite{truong-15}} - Exp.,
           $^{\rm m}$ Ref. {\cite{arimondo-77}} - Exp.,
           $^{\rm p}$ Ref. {\cite{he-20}} - Exp.,
           $^{\rm q}$ Ref. {\cite{yang-16}} - Exp.,
           $^{\rm r}$ Ref. {\cite{gilbert-83}} - Exp.,
           $^{\rm y}$ Ref. {\cite{chakraborty-23a}} - RCCSD,
           $^{\rm z}$ Ref. {\cite{roberts-23}} - AMPSCI,
           $^{\rm \beta}$ Ref. {\cite{dzuba-89}} - Relati. HF,
           $^{\rm \epsilon}$ Ref. {\cite{safronova-16}} - Relati. All-order,
           $^{\rm \zeta}$ Ref. {\cite{toh-19}} - Exp.,
           $^{\rm \theta}$ Ref. {\cite{damitz-19}} - Exp.,
           $^{\rm \kappa}$ Ref. {\cite {gregoire-15}} - Exp.,
           $^{\rm \lambda}$ Ref. {\cite{amini-03}} - Exp.,
           $^{\rm \nu}$ Ref. {\cite{patterson-15}} - Exp.,
   \end{flushleft}
   \label{tab_hfs}
   \end{center}
\end{table*}

We now discuss and analyze our results of magnetic HFS (MHFS) constants 
listed in Table \ref{tab_hfs}. Since $H_{\rm PNC}^{\rm NSD}$ depends on the
nuclear density, the wavefunction should be accurate within the nuclear region
to obtain reliable value of $E1^{\rm NSD}_{\rm PNC}$. Similarly, considering 
that MHFS operator has a Dirac-delta function, the wave function
should be accurate within the nuclear region to obtain accurate MHFS constants.
Thus, MHFS constants obtained from theoretical calculations serve as 
an important measure to indicate the accuracy of wavefunctions in the 
small $r$ region. For Cs there are several results of MHFS constants from 
previous works as it is a theoretically and experimentally well studied 
system. Our results along with those from the previous works are listed in 
Table \ref{tab_hfs}. Like in the case of excitation energies, to study the 
nature of electron correlations, we have provided the Breit and QED 
contributions separately. As evident from the table, for all the states our 
results for $A$ are in good agreement with the experimental values.
 
For the $6s_{1/2}$ state, our result of 2308 MHz is in good agreement with the 
experimental value 2298.2 MHz. The deviation of $\approx 0.5$\% is a trend 
common with the previous calculations. Among the theoretical results the 
smallest and largest deviations from experimental data are $\approx 0.2$\% 
\cite{kozlov-01b} and $\approx 0.8$\% \cite{derevianko-05}, respectively.
Compare to other theoretical results, ours is on the higher side. One reason
for this could be the inclusion of the corrections from the Breit and QED 
interactions. Without these our result of 2297.96 MHz is in better agreement
with the experimental value of 2298.2 MHz.

In the case of $6p_{1/2}$, our result of 287.8 MHz is lower than the 
experimental as well as other theoretical results. The deviation from the
average experimental value of 291.9 MHz is $\approx 1.4$\%. The contribution
from the Breit interaction is more than an order magnitude larger than the
QED correction and opposite in phase. As a result the Breit interaction reduces
the value of $A$. From the table it is evident that our result without the 
Breit and QED corrections is closer to the experimental result. It is 
$\approx 0.7$\% lower than the experimental data. An important observation is,
there is a variation among the previous theoretical results. The difference
between the lowest and highest values is $\approx 1.5$\%. There is limited
data for the $6p_{3/2}$ state. There are only two results each from the 
previous theoretical and experimental studies for comparison. Our result of 
49.79 MHz lies between the two experimental results, however, closer to the 
latest experimental result \cite{gerginov-03} with deviation of 
$\approx0.9$\%. Our computed value is in good agreement with the previous 
two theoretical results. 

Considering the state $7s_{1/2}$, there are several theoretical and 
experimental results from previous studies. Our result of 545.2 MHz is in 
excellent agreement with these data. It is $\approx 0.1$\% larger than the
average experimental value of 545.9 MHz. Again for the $7p_{1/2}$ state,
there are several theoretical results from previous studies, however, there
are only two experimental results. Most of the previous theoretical results 
are obtained using similar methods and have reported the similar values.
Our result of 93.0 MHz, is $\approx 1.3$\% smaller than the experimental 
result. However, like in the results of other states, our result without 
Breit and QED corrections is closer to the experimental value. Lastly, 
for $7p_{3/2}$ state, our result of 16.4 MHz is in good agreement with the 
experimental value and is $\approx 0.9$\% lower. Our result also matches well 
with the previous coupled-cluster result \cite{blundell-91}.

The $E1$ transition amplitudes from our calculations, along with the data from 
previous theoretical and experiment studies, are listed in the lower rows
of Table \ref{tab_hfs}. In contrast to MHFS constants, the results of $E1$ 
is a test for the accuracy of wavefunctions in large $r$ regions 
{\bf \cite{roberts-23}.} For the $E1$ amplitudes, there are several results 
from previous studies, and hence, a detailed comparison with each of the 
results is beyond the scope of the present work. However, as evident 
from the table, our results are in good agreement with the previous theory 
as well as experimental values listed in the table. In general, our results are 
closer to those reported in Refs. \cite{derevianko-05, porsev-10, 
chakraborty-23a}. This could be attributed to the similar CC based methods 
employed in these works. The small deviations from the previous results
could be due to the inclusion of corrections arising from the Breit and QED 
interactions, and the use of larger basis set in our calculations.


\subsection{Electric dipole polarizability}

In this section we discuss our results of $\alpha$ for the ground state 
$6s_{1/2}$, and excited states $6p_{1/2}$, $7s_{1/2}$ and $7p_{1/2}$. 
These states are chosen as these are expected to contribute dominantly 
to $E1^{\rm NSD}_{\rm PNC}$ through the strong PNC and dipole induced mixings. 
It is to be noted that, $\alpha$ of an atom or ion is an important property
relevant to several experiments to probe fundamental physics and technological
applications. This is applicable to the present work as well. In particular,
the computation of $\alpha$ involves the calculation of dipole matrix elements,
which also contribute to $E1^{\rm NSD}_{\rm PNC}$. To compute $\alpha$, we 
adopt two different methods. First, we employ the sum-over-states approach,
where we use $E1$ matrix elements obtained from our computations, and the
experimental energies to calculate $\alpha$ for $6s_{1/2}$, $6p_{1/2}$, 
$7s_{1/2}$ and $7p_{1/2}$ states. The values of $\alpha$ from this 
approach shall serve as a benchmark to gauge the accuracy of unperturbed 
wavefunctions and $E1$ transition amplitudes. And, second we use the FS-PRCC 
method, discussed earlier in the context of PNC perturbation, to compute 
$\alpha$ for $6s_{1/2}$ and $7s_{1/2}$ states. The values of $\alpha$ from 
this shall serve as an indicator of accuracy for the initial and final PNC 
perturbed states $\widetilde{6s}_{1/2}$ and $\widetilde{7s}_{1/2}$. 
Since the Goldstone diagrams for the polarizability are topologically same 
as PNC diagrams in the electronic sector, obtaining accurate value of  
$\alpha$ using FS-PRCC shall lead to accurate value of $E1^{\rm NSD}_{\rm PNC}$.

\begin{table}[!ht]
	\caption{Electric dipole polarizability (a.u.) of $6s_{1/2}$ and $6p_{1/2}$ states
	of Cs using sum-over-states approach. The dominant contributing 
	E1 matrix elements are provided to quantify the nature of electron 
	correlations.}
  \begin{ruledtabular}
  \begin{tabular}{lr}
   \text{Contr./E1 matrix} & $\alpha_{6s_{1/2}}$  \\
       	\hline
   $6p_{3/2} \rightarrow 6s_{1/2}$ &  255.73 \\
   $6p_{1/2} \rightarrow 6s_{1/2}$ &  134.77 \\
   $7p_{3/2} \rightarrow 6s_{1/2}$ &  1.23   \\
   $7p_{1/2} \rightarrow 6s_{1/2}$ &  0.28   \\
   $8p_{3/2} \rightarrow 6s_{1/2}$ &  0.22   \\
   Core  		        &  15.8   \\
   Total                        &  408.03          \\
   Others          & $396.02^a$, $399.9(1.9)^b$, $401.0(6)^c$ \\
                                & $401.5^d$, $398.4(19)^e$ \\
   Exp.            & $402(8)^f$, $401.0(6)^g$  \\
       	\hline
   \text{Contr./E1 matrix} & $\alpha_{6p_{1/2}}$  \\
       	\hline
   $5d_{3/2} \rightarrow 6p_{1/2}$   & 1215.98     \\
   $7s_{1/2} \rightarrow 6p_{1/2}$   & 180.56\\
   $6s_{1/2} \rightarrow 6p_{1/2}$  & -134.76\\
   $6d_{3/2} \rightarrow 6p_{1/2}$   & 89.45  \\
   $7d_{3/2} \rightarrow 6p_{1/2}$   & 34.39  \\
   $8d_{3/2} \rightarrow 6p_{1/2}$   & 21.45  \\
   $8s_{1/2} \rightarrow 6p_{1/2}$   & 6.62\\
   $9d_{3/2} \rightarrow 6p_{1/2}$   & 6.54  \\
   $9s_{1/2} \rightarrow 6p_{1/2}$   & 3.64\\
   $10s_{1/2} \rightarrow 6p_{1/2}$  & 1.70     \\
   $11s_{1/2} \rightarrow 6p_{1/2}$  & 0.50    \\
   Core                              & 15.8 \\
   Total                             & 1441.87 \\
   Other cal.           & $1338(5.4)^e$, $1404(28)^h$, $1327^i$ \\
   Exp.  		& $1371^j$, $1328.35^k$ \\
  \end{tabular}
  \end{ruledtabular}
  \begin{flushleft}
   $^{\rm a}$ Ref. \cite{lim-05} - RCCSDT,
   $^{\rm b}$ Ref. \cite{safronova-99} - RCCSD,
   $^{\rm c}$ Ref. \cite{chakraborty-23} - sum-over-states,
   $^{\rm d}$ Ref. \cite{derevianko-99b} - RLCCSD,
   $^{\rm e}$ Ref. \cite{tchoukova-07}- RLCCSDT, \\
   $^{\rm f}$ Ref. \cite{molof-74} - Exp.,
   $^{\rm g}$ Ref. \cite{amini-03} - Exp.,
   $^{\rm h}$ Ref. \cite{safronova-04} - sum-over-states,\\
   $^{\rm i}$ Ref. \cite{zhou-89} - RHF,
   $^{\rm j}$ Ref. \cite{hunter-88} - Exp.,
   $^{\rm k}$ Ref. \cite{hunter-92} - Exp.,
  \end{flushleft}
  \label{pol_sum1}
\end{table}

\begin{table}[!ht]
	\caption{Electric dipole polarizability (a.u.) of $7s_{1/2}$ 
	and $7p_{1/2}$ states of Cs using sum-over-states approach. 
	The dominant contributing E1 matrix elements are provided to 
	quantify the nature of electron correlations.}
  \begin{ruledtabular}
  \begin{tabular}{lr}
   \text{Contr./E1 matrix} & $\alpha_{7s_{1/2}}$ \\
   \hline
   $7p_{3/2} \rightarrow 7s_{1/2}$  & 4417.09 \\
   $7p_{1/2} \rightarrow 7s_{1/2}$  & 2414.83 \\
   $6p_{3/2} \rightarrow 7s_{1/2}$  & -451.13 \\
   $6p_{1/2} \rightarrow 7s_{1/2}$  & -180.56 \\
   $8p_{3/2} \rightarrow 7s_{1/2}$  &  33.50  \\
   $8p_{1/2} \rightarrow 7s_{1/2}$  &  10.25  \\
   $9p_{3/2} \rightarrow 7s_{1/2}$  &  2.50    \\
   $9p_{1/2} \rightarrow 7s_{1/2}$  &  0.35    \\
   $10p_{1/2} \rightarrow 7s_{1/2}$ & 0.01    \\
   $10p_{3/2} \rightarrow 7s_{1/2}$ & 0.00    \\
  Core  				            & 15.8     \\
  Total                             & 6256.64  \\
  Other cal.                        & $6238^a$, $6061^b$, $6140^c$ \\
  Exp.                              & $6238^d$\\
  \hline
  \text{Contr./E1 matrix} & $\alpha_{7p_{1/2}}$ \\
  \hline
  $6d_{3/2} \rightarrow 7p_{1/2}$   & 29919.87 \\
  $8s_{1/2} \rightarrow 7p_{1/2}$   & 2570.51 \\
  $7s_{1/2} \rightarrow 7p_{1/2}$   & -2413.06 \\
  $7d_{3/2} \rightarrow 7p_{1/2}$   & 890.96 \\
  $8d_{3/2} \rightarrow 7p_{1/2}$   & 83.87 \\
  $5d_{3/2} \rightarrow 7p_{1/2}$   & -65.39 \\
  $9s_{1/2} \rightarrow 7p_{1/2}$   &  52.75 \\
  $9d_{3/2} \rightarrow 7p_{1/2}$   & 22.26 \\
  $10s_{1/2} \rightarrow 7p_{1/2}$  & 6.72 \\
  Core                              & 15.8   \\
  Total                             & 31084.29\\
  Other cal.                        & $29900(700)^a$\\
  Exp.                              & $29600(600)^e$
  \end{tabular}
  \end{ruledtabular}
  \begin{flushleft}
   $^{\rm a}$ Ref. \cite{tchoukova-07}- RLCCSDT,
   $^{\rm b}$ Ref. \cite{zhou-89} - RHF,
   $^{\rm c}$ Ref. \cite{van-94} - sum-over-states.,
   $^{\rm d}$ Derived from the experiments Refs. \cite{bennett-99} and \cite{amini-03},
   $^{\rm e}$ Ref. \cite{domelunksen-83} - Exp.
  \end{flushleft}
  \label{pol_sum2}
\end{table}

\begin{figure}
\begin{center}
  \includegraphics[scale = 0.35, angle = -90]{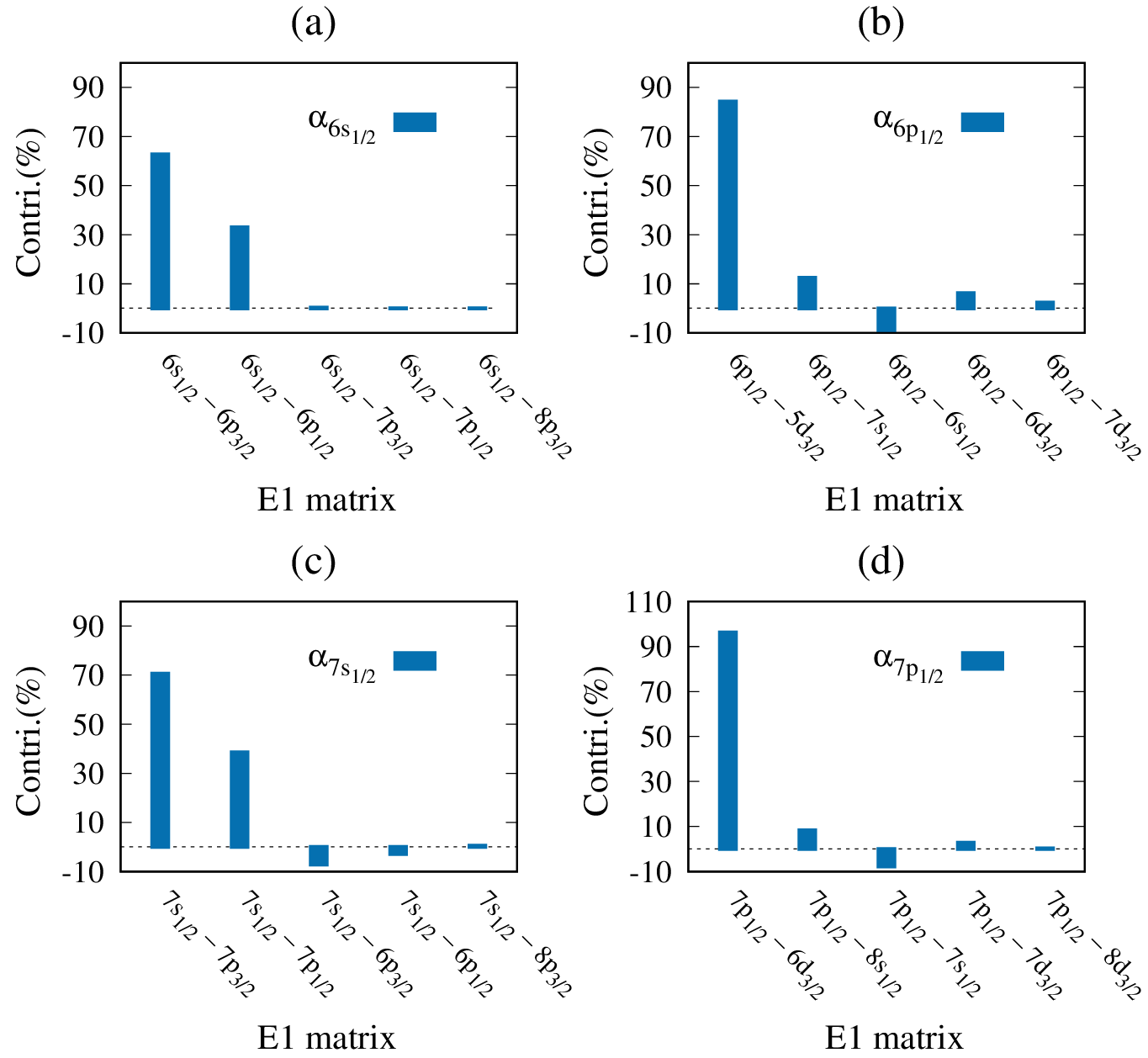}
  \caption{Five dominant contributing dipolar mixings to $\alpha$ for 
	$6s_{1/2}$, $6p_{1/2}$, $7s_{1/2}$, and $7p_{1/2}$ states.}
\label{pol_orb}
\end{center}
\end{figure}

In the Tables \ref{pol_sum1} and \ref{pol_sum2}, we have listed our results 
of $\alpha$ computed using sum-over states approach for $6s_{1/2}$, 
$6p_{1/2}$ and $7s_{1/2}$, $7p_{1/2}$ states, respectively. For comparison, 
we have also listed the results from other theoretical and experimental
works. To assess the trend of electron correlations, contributions from 
the dominant $E1$ matrix elements are listed separately. For $6s_{1/2}$ state, 
as can be expected, the dominant contribution comes from the dipole mixing 
with $6p_{1/2}$ and $6p_{3/2}$ states. As discernible from Fig. \ref{pol_orb}, 
they contribute $\approx 33$\% and $63$\%, respectively, to the total value. 
Contribution from the core polarization, $\approx 4$\%, is the next leading 
order contribution. Comparing with the results from previous works, our value 
is larger than the other theoretical and experimental results. Our value 
of 408.03 is $\approx 1.7$\% larger than the recent experiment \cite{amini-03}. 
The reason for this could be attributed to larger values of 
$6s_{1/2}\rightarrow 6p_{1/2}$ and $6s_{1/2}\rightarrow 6p_{3/2}$ matrix 
elements than experiment. Contributions from the intermediate states 
with $n \ge 8$ are found to be small.    

For the $6p_{1/2}$ state, the dominant contribution of $\approx 84$\% 
arises from the dipole mixing with $5d_{3/2}$ state. The next two 
dominant contributions are from the $7s_{1/2}$ and $6s_{1/2}$ states. These 
contribute $\approx 12$\% and $-9$\%, respectively. The other two significant 
contributions are from the $6d_{3/2}$ and $7d_{3/2}$ states contributing
$\approx 6$\% and $2$\%, respectively. Comparing with the previous theoretical 
works, there is a spread in the reported values. The result from the
relativistic Hartree-Fock method in Ref. \cite{zhou-89} is $\approx 6$\% 
smaller than the sum-over-states value in Ref. \cite{safronova-04}. A similar
trend is also there among the experimental values. Our result is 
closer to the sum-over-states value \cite{safronova-04}. 

For the $7s_{1/2}$ state, two states $7p_{1/2}$ and $7p_{3/2}$ contribute 
dominantly to $\alpha$, their combined result 6831.92 a.u. is larger than 
the net value 6256.64 a.u. The reason is, the contributions from 
both the next two dominant states $6p_{1/2}$ and $6p_{3/2}$ are in opposite 
phase. These two states together has a contribution of $-631.69$ a.u. The
contributions from the remaining intermediate states $np_{1/2}$ and $np_{3/2}$
continue to decrease with higher $n$. We incorporate upto $n=10$ and 
contribution from $10p_{3/2}$ is negligible and hence, contributions from 
$n>10$ can be neglected. Our net value, 6256.64 a.u. is in good agreement 
with the experimental value, 6238 a.u. Considering the previous calculations, 
there are large differences among the reported values, 6061 \cite{zhou-89}, 
6140 \cite{van-94} and 6238 \cite{tchoukova-07}.  Our result is on the 
higher side of these and closer to Ref. \cite{tchoukova-07}. For the 
$\alpha$ of $7p_{1/2}$, state $6d_{3/2}$ has the leading contribution
$\approx 96$\%. The next leading contributions, almost equal in magnitude
but opposite in phase, arise from the $7s_{1/2}$ and $8s_{1/2}$ states. 
These states together contribute $\approx 8$\%. Like the trends of other 
states, our sum-over-states results are higher than
the previous theoretical results. It should, however, be noted that, the 
$E1$ matrix elements used in the calculations of $\alpha$ are all from our 
FSRCC calculations, which is {\em ab initio}.

\begin{table}
\caption{Electric dipole polarizability (a.u.) of the ground state, $6s_{1/2}$, and
        excited state $7s_{1/2}$ of Cs using FS-PRCC theory. For quantifying 
	electron correlations embedded in FS-PRCC, the termwise contributions 
	are listed separately.}
\begin{ruledtabular}
    \begin{tabular}{lrr}
            Term + H. c.                    &$6s_{1/2}$ & $7s_{1/2}$   \\
        \hline
        Dirac-Fock                          &   647.257 & 8060.598      \\
	    ${\mathbf D}{\mathbf S}^{(1)}_{1}$                    &  -188.874 & -1423.122    \\
	    ${\mathbf D}{\mathbf S}^{(1)}_{2}$            &  -14.867  &  -35.864    \\
	    ${S^{(0)}_{1}}^\dagger {\mathbf D} {\mathbf S}^{(1)}_1$ &  -23.789  &  -338.479    \\
	    ${S^{(0)}_{2}}^\dagger {\mathbf D} {\mathbf S}^{(1)}_1$ &  -13.816  &  -29.974   \\
	    ${S^{(0)}_{2}}^\dagger {\mathbf D} {\mathbf S}^{(1)}_2$ &  -0.702   &  -1.641     \\
	    ${ \mathbf D}{\mathbf T}^{(1)}$                        &  -0.722   &  -0.120    \\
	    ${T^{(0)}}^\dagger {\mathbf D} {\mathbf T}^{(1)}$       &  -0.002   &  0.001    \\
	    ${T^{(0)}}^\dagger {\mathbf D} {\mathbf S}^{(1)}$       &   1.929   &  4.818   \\
	    ${S^{(0)}}^\dagger {\mathbf D} {\mathbf T}^{(1)}$       &  -1.570   &  -0.649   \\
        Normalization                       &  -4.773   &  -11.437   \\
	    Total PRCC                          &  400.071  &  6224.131  \\
         PRCC + Breit                       &  400.107  &  6224.691 \\
         PRCC + Breit + QED                 &  399.895  &  6221.392 \\
         PRCC + Breit + QED + Triples       &  399.901  &  6221.423 \\
	 Recommended       &  399.90$\pm$ 4.0  &  6221.42$\pm$ 62.2\\
    \end{tabular}
\end{ruledtabular}
\label{pol_prcc}
\end{table}

In the Table \ref{pol_prcc}, we have listed the FS-PRCC results of $\alpha$ for 
initial and final PNC-perturbed states, $6s_{1/2}$ and $7s_{1/2}$. These
computations are done using a  basis set of moderate 
size ($11s$, $10p$, $8d$, $10f$, $8g$, $8h$). 
To improve the results further, corrections from the Breit, QED and perturbative 
triples are also included. As it is evident from the table, our PRCC results 
of 399.90 a.u. and 6221.42 a.u. for the $6s_{1/2}$ and $7s_{1/2}$ states, 
respectively, are in excellent agreement with the experimental results
401 a.u. \cite{amini-03} and 6238 a.u. \cite{bennett-99, amini-03}, with a 
small deviation of $\approx 0.3$\% in each case. The reason for such 
a good match could be attributed to the merits of the FS-PRCC theory to 
incorporate the effect of external perturbation in many-body calculations. 
In contrast to the sum-over-states approach, the FS-PRCC theory subsumes the 
effects of all intermediate states implicitly through the PRCC operator.

Examining the termwise contributions, as expected, in both the states the 
dominant contribution arises from the Dirac-Fock (DF). The next 
leading order contribution is from ${\mathbf D}{\mathbf S}^{(1)}_{1}$ + H.c, 
however it is opposite in phase to the dominant term. So, there is a 
significant cancellation between the two important contributions. 
A large contribution from this term is expected as 
${\mathbf D}{\mathbf S}^{(1)}_{1}$ subsumes the dominant contributions 
from core polarization and {\em valence-virtual} correlation effects.
The next two leading order terms are
${S^{(0)}_{1}}^\dagger {\mathbf D}{\mathbf  S}^{(1)}_1$ + H.c. and 
${\mathbf D}{\mathbf S}^{(1)}_{2}$ + H.c. They contribute $\approx -6(-6)$\%
and $-4(0.6)$\%, respectively, for $6s_{1/2}$($7s_{1/2}$) state. Further,
the terms ${S^{(0)}_{2}}^\dagger {\mathbf D}{\mathbf S}^{(1)}_1$ + H.c., 
which are second order in cluster operators, have contributions of 
similar order to ${\mathbf D{\mathbf S}}^{(1)}_{2}$. Other terms with two
orders of CC operators have small contributions. 

\subsection{NSD-PNC amplitudes}

In this section we present and discuss the results of 
$E1^{\rm NSD}_{\rm PNC}$, which are listed in Table \ref{nsd_tab} for 
transitions between different hyperfine states. To quantify the importance of
electron correlations, in the table we have listed results from different 
methods with increasing sophistication. Thus, among the methods, DF listed 
as first does not include any electron-correlation and the last,
PRCC(T)+Bre.+QED is the one which includes maximum electron-correlation 
effects among all the methods. As expected, the dominant contribution 
to $E1^{\rm NSD}_{\rm PNC}$ is from the DF for all the hyperfine transitions. 
It contributes $\approx 91$\%, $87$\%, $84$\% and $91$\% of the PRCC value 
for the $3\rightarrow3$, $3\rightarrow4$, $4\rightarrow3$ and $4\rightarrow4$ 
hyperfine transitions, respectively. The next leading order contribution is 
from the first-order MBPT, where contribution of $\approx -13.5$\% is observed 
for both $3\rightarrow3$ and $4\rightarrow4$ transitions. Whereas, there are
in-phase contributions of $\approx 16.5$\% and 12.2\% for $3\rightarrow4$ 
and $4\rightarrow3$ transitions, respectively. Similarly, the second and 
third-order MBPT also show a mix trend of opposite and in-phase contributions, 
however, with smaller magnitudes. Among the four transitions the contribution
from MBPT(2) is larger ($\approx -2$\% ) in the case of $3\rightarrow4$ and 
$4\rightarrow3$. And, MBPT(3) is, however, has a more prominent 
contribution of $\approx 3$\% in the case of $3\rightarrow3$ and 
$4\rightarrow4$ transitions.

\begin{table}
        \caption{The $E1^{\rm NSD}_{\rm PNC}$ of $6s_{1/2}\rightarrow 7s_{1/2}$ 
	transition in units of $iea_{0} \times 10^{-12} \mu^{'}_{W} $. To 
        assess the nature of electron correlations, contributions from 
        different levels of the theory used is provided separately. For 
        comparison, the data available from other theory calculations are 
        also provided.}
        \label{orbe_tl}
        \begin{ruledtabular}
        \begin{tabular}{lcccc}
        \text{Methods} &  \multicolumn{4}{c}{Transition $F_v \rightarrow F_w$} \\
        \hline
         &  $3\rightarrow 3$ & $3\rightarrow 4$ & $4\rightarrow 3$ & $4\rightarrow 4$ \\
        \hline
         DF & 1.821 & 4.729 & 5.430 & 2.073    \\
       	 MBPT(1) & 1.551  & 5.624 & 6.220 & 1.766 \\
         MBPT(2) & 1.570  & 5.492 & 6.097 & 1.788  \\
         MBPT(3) & 1.636  & 5.490 & 6.120 & 1.863    \\
         PRCC    & 1.997  & 5.419 & 6.452 & 2.273    \\
         PRCC(T) &  1.998 & 5.420 & 6.453 & 2.273   \\
	 PRCC(T)+Bre.    & 1.993   & 5.406  &6.368  & 2.268   \\
	 PRCC(T)+Bre.    & 2.000   & 5.425  &6.250  & 2.276    \\
	 + QED           &         &        &       &           \\
	 Reco.           & 2.00$\pm$0.02 & 5.43$\pm$0.05 & 6.25$\pm$0.06 & 2.28$\pm$0.02 \\
         Other cal & $2.249^{\rm a}$  & $6.432^{\rm a}$    & $7.299^{\rm a}$ & $2.560^{\rm a}$ \\ 
		   & $2.274^{\rm c}$  & $7.948^{\rm b}$   & $7.057^{\rm b}$ & $2.589^{\rm c}$  \\                            
	           & $2.334^{\rm d}$  & $5.446^{\rm c}$,$6.496^{\rm d}$ & $6.313^{\rm c}$,$7.394^{\rm d}$&$2.658^{\rm d}$  \\            
        \end{tabular}
        \end{ruledtabular}
\begin{flushleft}
      $^{\rm a}$ Ref. {\cite{johnson-03}},
      $^{\rm b}$ Ref. {\cite{safronova-09a}}
      $^{\rm c}$ Ref. {\cite{mani-11cs}}
      $^{\rm d}$ Ref. {\cite{chakraborty-24}}
\end{flushleft}
\label{nsd_tab}
\end{table}

\begin{table}
\caption{Termwise contributions to $E1^{\rm NSD}_{\rm PNC}$. Values listed 
	are in the unit of $iea_{0} \times 10^{-12} \mu^{'}_{W} $}
\begin{ruledtabular}
\begin{tabular}{lcccccccc}
 Term + H. c.   & $3 \rightarrow 3$ & $3 \rightarrow 4$ & $4 \rightarrow 3$ & $4 \rightarrow 4$ \\
\hline
 $\mathbf D{\mathbf S}^{(1)}_{1}$                         &  3.3696 & 6.3157  & 7.9280 & 3.8634\\
 $\mathbf D{\mathbf S}^{(1)}_{2}$                         & -0.0643 & -0.3453 &-0.4078 & -0.0732\\
 ${S^{(0)}_{1}}^\dagger \mathbf D {\mathbf S^{(1)}_{1}}$  & -1.0649 & -0.3784 &-0.7988 & -1.2124\\
 ${S^{(0)}_{2}}^\dagger \mathbf D {\mathbf S^{(1)}_{1}}$  & -0.0653 & 0.1455  & 0.1288 & -0.0744\\
 ${S^{(0)}_{2}}^\dagger \mathbf D {\mathbf S^{(1)}_{2}}$  & -0.0023 & -0.0271 &-0.0337 & -0.0027\\
 $\mathbf D{\mathbf T}^{(1)}$                             & -0.0993 & -0.1936 &-0.2318 & -0.1130\\
 ${T^{(0)}_{1}}^\dagger \mathbf D {\mathbf T^{(1)}_{1}}$  & 0.0000  & -0.0028 &-0.0028 & 0.0000\\
 ${T^{(0)}_{2}}^\dagger \mathbf D {\mathbf T^{(1)}_{1}}$  & -0.0131 & 0.0074  & 0.0024 & -0.0149\\
 ${S^{(0)}_{1}}^\dagger \mathbf D {\mathbf T^{(1)}_{1}}$  & -0.0083 & 0.0273  & 0.0241 & -0.0094\\
 ${S^{(0)}_{2}}^\dagger \mathbf D {\mathbf T^{(1)}_{1}}$  & -0.0132 & -0.0213 &-0.0264 & -0.0150 \\
 ${T^{(0)}_{2}}^\dagger \mathbf D {\mathbf T^{(1)}_{2}}$  & -0.0031 & -0.0020 & -0.0032 & -0.0035   \\
Normalization                       & -0.0392 & -0.1063 & -0.1266 & -0.0446  \\
     Total                                &  1.9966 &  5.4190  & 6.4521 & 2.2732 \\
\end{tabular}
\end{ruledtabular}
\label{nsd_termwise}
\end{table}

Among the previous theoretical works, we could find two published works
which reported the values of ${E1}^{\rm NSD}_{\rm PNC}$ for all the four 
transitions. The first one is by Johnson and collaborators using 
RPA \cite{johnson-03}, and second is by Safronova and collaborators 
\cite{safronova-09a} using the all-order method. For $3\rightarrow3$ and 
$4\rightarrow4$ transitions, our results of $2.0 \times 10^{-12}$ and 
$2.28 \times 10^{-12}$, respectively, are lower than the RPA results. The 
reason for this may be attributed to improved consideration of electron 
correlations in our work. In our computations, the residual Coulomb 
interaction is incorporated to all orders of perturbation 
using relativistic FSRCC theory. Similarly, the PNC interaction is treated 
using the FS-PRCC theory. As discussed earlier, in contrast to the
sum-over-states approach, PRCC includes contributions from all the 
possible intermediate states in the calculation of ${E1}^{\rm NSD}_{\rm PNC}$.

For the $3\rightarrow4$ and $4\rightarrow3$ transitions, there is a large 
difference between the previous results. For example, the all-order 
value \cite{safronova-09a} for $3\rightarrow4$ transition is $\approx 20$\% 
higher than the RPA value \cite{johnson-03}. This implies that the electron 
correlation effects must be treated accurately to get reliable values of 
${E1}^{\rm NSD}_{\rm PNC}$. Our results are lower than both the previous 
results. It is, however, to be noted that our DF results of 
$1.82 \times 10^{-12}$, $4.73 \times 10^{-12}$, 
$5.43 \times 10^{-12}$ and $2.07 \times 10^{-12}$, for 
$3\rightarrow3$, $3\rightarrow4$, $4\rightarrow3$ and $4\rightarrow4$ 
transitions, respectively, are in good agreement with the values 
reported in Ref. \cite{johnson-03}. Thus, the difference from the previous
works \cite{johnson-03, safronova-09a} is due to better consideration of 
electron correlations in our work. 

Apart from Refs. \cite{johnson-03, safronova-09a}, there are two 
previous works \cite{mani-11cs, chakraborty-24} which report results 
on $E1^{\rm NSD}_{\rm PNC}$ for considered hyperfine transitions.
Both of the works use CC method like the present work, however, with some key 
differences. There is a slight difference in the $E1^{\rm NSD}_{\rm PNC}$ 
values with respect to Ref. \cite{mani-11cs} for all the hyperfine transitions. 
This could be attributed to improved consideration of electron correlation
effects due to the inclusion of nonlinear terms of the PRCC theory 
in the present work. Whereas Ref. \cite{mani-11cs} is based on 
{\em linearized} PRCC theory. Our results of $E1^{\rm NSD}_{\rm PNC}$ 
are lower than Ref. \cite{chakraborty-24} for all hyperfine 
transitions. One possible reason for this could be the incorporation 
of the corrections from Breit interaction, QED effects and perturbative 
triples. These were not considered in the previous works. 
Thus, the present work includes important many-body effects not 
included in the previous theoretical studies of NSD-PNC of Cs.

The Table \ref{nsd_termwise} lists the term wise contribution to 
${E1}^{\rm NSD}_{\rm PNC}$. As seen from the table, for all the hyperfine 
transitions, the leading order (LO) contribution is from the 
$\mathbf{D}{\mathbf{S}}^{(1)}_{1}$ + H.c. term. This is natural, as it
subsumes the contribution from DF, and the dominant contributions from 
{\em core-polarization} and {\em valence-virtual} correlation effects.  
It contributes $\approx 169$ and $170$\%, of the total PNC 
amplitude, for $3\rightarrow3$ and $4\rightarrow4$ transitions, respectively. 
Whereas, the contribution is $\approx 117$ and $123$\% for $3\rightarrow4$ 
and $4\rightarrow3$ transitions, respectively. The next leading order (NLO) term is 
${S^{(0)}_{1}}^\dagger \mathbf{D}{\mathbf{S}^{(1)}_{1}}$ + H.c.
It has an opposite contribution of $\approx -53$\% for both the 
diagonal transitions. For off-diagonal transition $3\rightarrow4$ and 
$4\rightarrow3$, however, contributions vary, to $\approx -7$ 
and $-12$\%, respectively.
Based on the LO and NLO contributions one can infer that the one-body perturbed 
cluster operator, $\mathbf{S}^{(1)}_{1}$, subsumes the dominant correlation 
effects from the PNC perturbation. For the next dominant contribution, 
the diagonal and off-diagonal hyperfine transitions show different 
trends of electron correlation effects. Interestingly, unlike in $\alpha$, 
for $3\rightarrow3$ and $4\rightarrow4$ transitions, the next dominant 
contribution arise from the core electrons through 
the $\mathbf{D}\mathbf{T}^{(1)}$ + h. c. terms. It contributes $\approx -5$\% 
of the total amplitude. However, for the $3\rightarrow4$ and 
$4\rightarrow3$ transitions the term $\mathbf{D}{\mathbf{S}}^{(1)}_{2}$ + H.c. 
provides the dominant contribution of $\approx -6$\% for both the transitions.

\begin{figure}
\begin{center}
  \includegraphics[scale = 0.35, angle = -90]{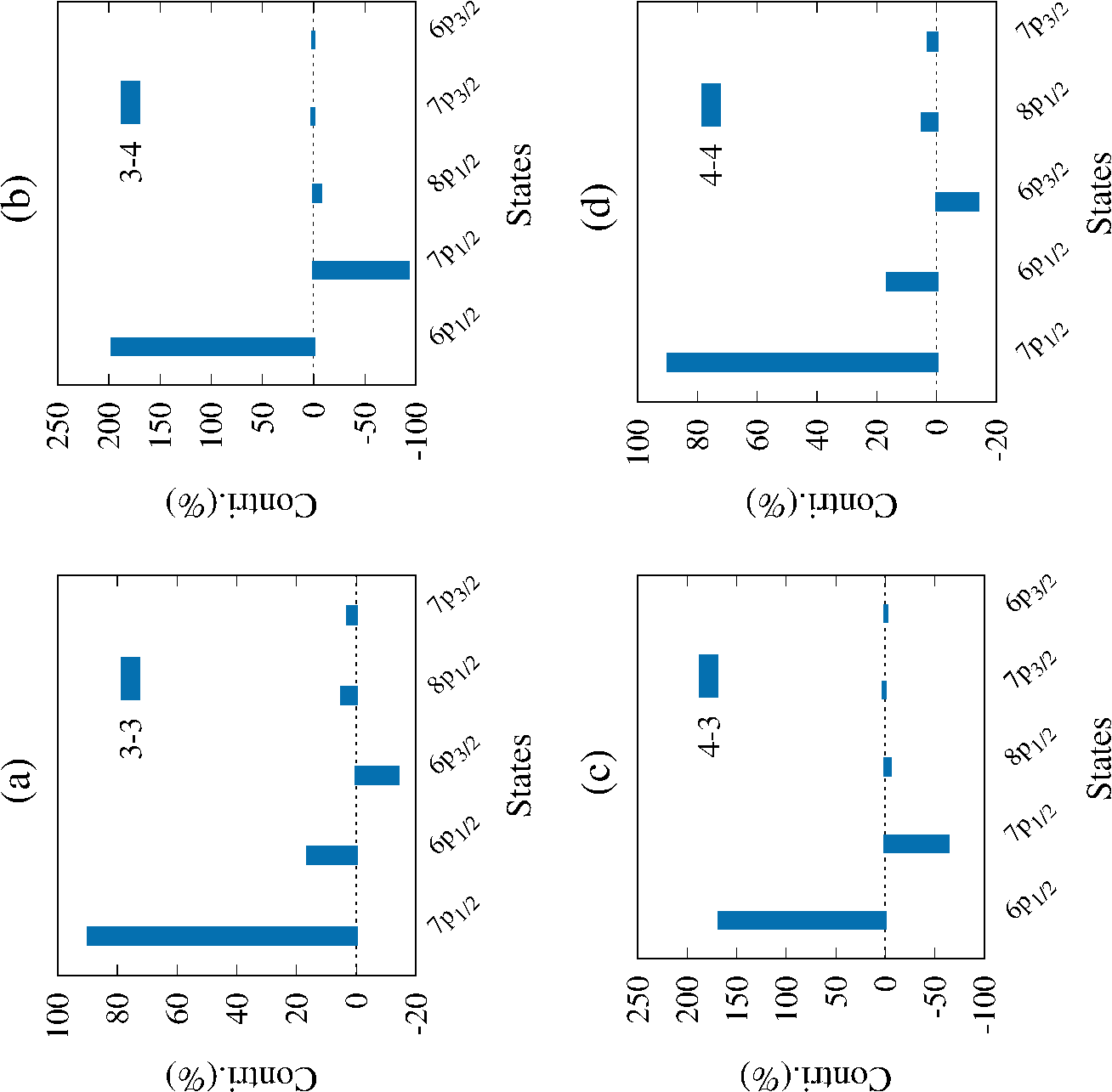}
  \caption{Five dominant contributions from orbitals to $E1^{\rm NSD}_{\rm PNC}$. 
	The percentage contribution is calculated with respect to the 
	leading order term $\mathbf{D}\mathbf{S}^{(1)}_{1}$.}
  \label{fig_orbital}
\end{center}
\end{figure}

\subsection{PNC mixing, correlation effects, and other corrections}

  To examine the different many-body effects in finer detail, we consider
and analyze the important terms. To begin with we take the PNC induced 
orbital mixing. This may be discernible from the orbital-wise contributions to
the LO term $\mathbf{D}\mathbf{S}^{(1)}_{1}$. As seen from 
Fig. \ref{fig_orbital}, the $3\rightarrow3$ and $4\rightarrow4$ transitions
have the same trend. In both, interestingly, the largest contribution 
$\approx 90$\% of $\mathbf{D}\mathbf{S}^{(1)}_{1}$ arises from the PNC 
induced mixing with $7p_{1/2}$ state. This is due to small energy difference 
with $7s_{1/2}$ state. The next dominant mixing is from the $6p$ state. 
In particular, $6p_{1/2}$ state has a contribution of $\approx 16$\%,
whereas the $6p_{3/2}$ state has an opposite phase contribution of 
$-14$\%. The next two significant contributions are from the $8p_{1/2}$ and 
$7p_{3/2}$ states, they contribute $\approx 5$\% and 3\%, respectively. 
The $3\rightarrow4$ and $4\rightarrow3$ transitions also have a similar trend. 
For both the transitions, the dominant PNC mixing is with the $6p_{1/2}$ state. 
Contributions of $\approx 196$ and $167$\% to $\mathbf{D}\mathbf{S}^{(1)}_{1}$ 
is observed for the $3\rightarrow4$ and $4\rightarrow3$ transitions, respectively.
The next largest mixing with an opposite phase of $\approx -92$\% and 
$-63$\%, respectively, are there from the $7p_{1/2}$ state. 
The $8p_{1/2}$ and $7p_{3/2}$ are the next two dominant contributers. There is
a contribution of $\approx -7 (1.2)$\% and $\approx -5$\%(1.5\%) for the
$3\rightarrow4$ and $4\rightarrow3$ transitions, respectively, from the
$8p_{1/2} (7p_{3/2})$ state. The $6p_{3/2}$ state is the fifth dominant
contribution with $\approx 1.0$\% and $-1.4$\%, respectively, to the
two transitions. The other important contributions which merit closer
study are the core-polarization effects, valence-virtual correlation, 
Breit and QED effects, and triple corrections. Each of these are addressed
in detail.

\begin{figure}
\begin{center}
  \includegraphics[scale = 0.8]{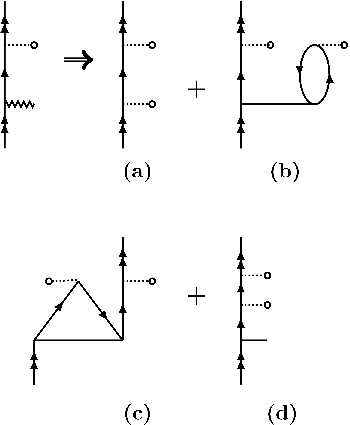}
	\caption{The DF (diagram (a)), core-polarization (diagrams (b) and (c)) 
	and valence-virtual (diagram (d)) contributions subsumed in the 
	term $\mathbf{D}\mathbf{S}_1^{(1)}$.} 
  \label{fig_cp}
\end{center}
\end{figure}


\subsubsection{Core polarization and valence-virtual correlation effects}

\begin{figure}
\begin{center}
  \includegraphics[scale = 0.35, angle = -90]{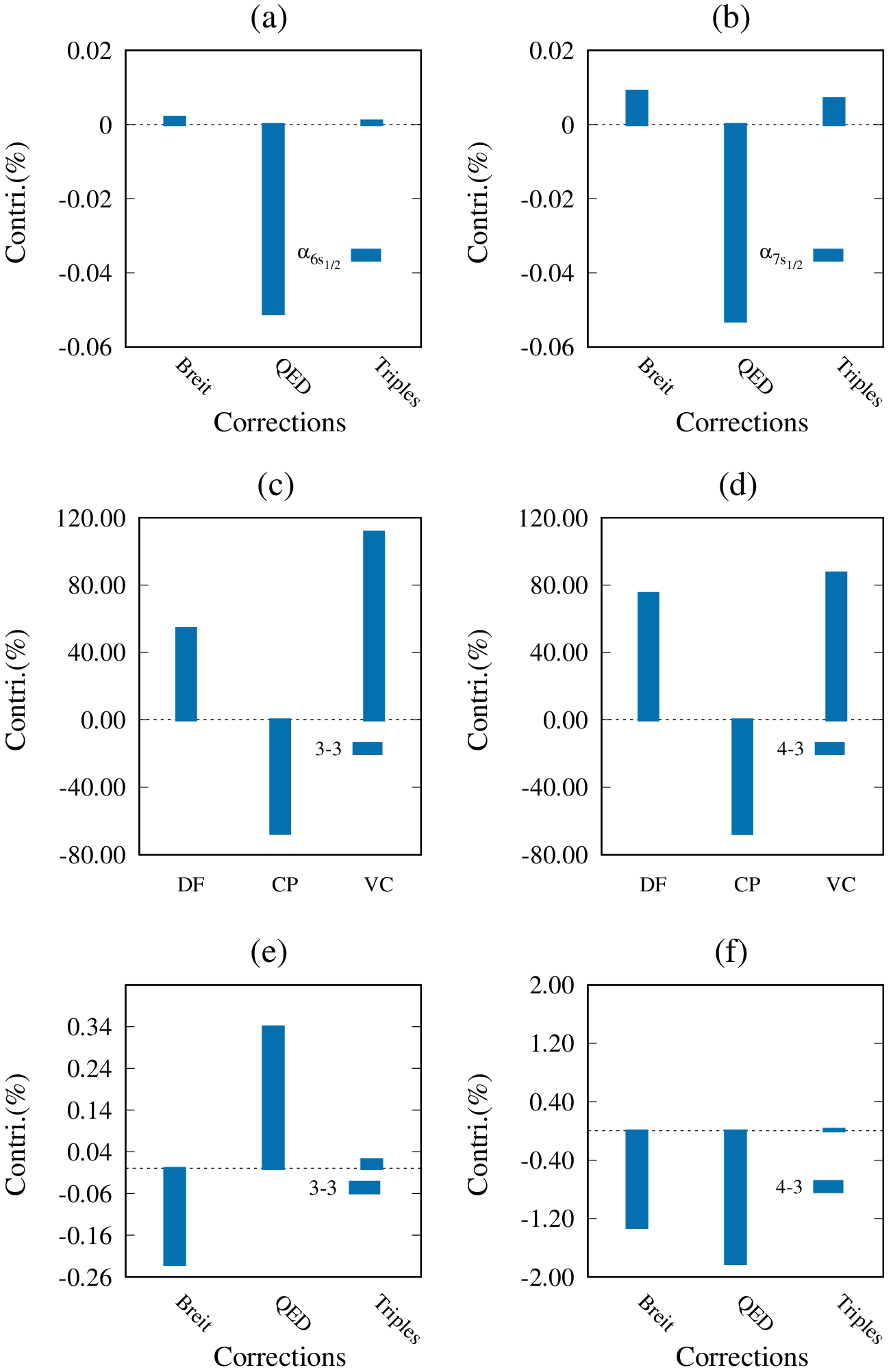}
  \caption{The percentage contributions from Breit, QED and perturbative
	triples to $\alpha$ (panels (a), (b)) and 
	${E1}^{\rm NSD}_{\rm PNC}$ (panels (e), (f)). The percentage
	contributions from DF, CP and VC to ${E1}^{\rm NSD}_{\rm PNC}$ 
	(panels (c), (d)).}
  \label{fig_qed}
\end{center}
\end{figure}

To quantify the contributions from core polarization and {\em valence-virtual}
correlation effects to ${E1}^{\rm NSD}_{\rm PNC}$, we again consider
the LO term $\mathbf{D}\mathbf{S}_1^{(1)}$. As discernible from the 
Goldstone diagrams in Fig. \ref{fig_cp}, the term 
$\mathbf{D}\mathbf{S}_1^{(1)}$ subsumes the contributions from DF, 
{\em core-polarization} (CP) and {\em valence-virtual} correlation (VC) 
effects. In particular, the diagrams Fig. \ref{fig_cp}(b) and 
\ref{fig_cp}(c) contribute to CP through dipole mixing between unperturbed 
states. The other dominant contribution to CP is from $\mathbf{S}_2^{(1)}$
through the term $\mathbf{D}\mathbf{S}_2^{(1)}$. And, the corresponding
diagrams are Fig. \ref{e1pnc_cc_fig}(b) and its exchange.  An estimate of 
the VC correlation is obtained by subtracting the contributions of the first 
three diagrams in Fig. \ref{fig_cp} from the $\mathbf{D}\mathbf{S}^{(1)}$ 
value. In Fig. \ref{fig_qed} we have shown the contributions from DF, CP 
and VC to the $3\rightarrow3$ and $4\rightarrow3$ hyperfine transitions.
As discernible from Fig. \ref{fig_qed} (c) and \ref{fig_qed}(d), as expected, 
DF has the largest contribution. It is $\approx 91$\% and 84\% of
total value to $3\rightarrow3$ and $4\rightarrow3$ transitions, respectively.
Considering the CP and VC contributions, they are in opposite phase for 
all the hyperfine transitions, and in terms of magnitude, VC has a larger
contribution. For example, for $3\rightarrow3$ diagonal transition, VC 
contributes $\approx 111$\% to the value of $\mathbf{D}\mathbf{S}_1^{(1)}$, 
whereas the contribution from CP is $\approx -67$\%. A similar trend of 
contributions with $\approx 87$\% and $\approx -67$\%, respectively, is also 
observed for the off-diagonal $4\rightarrow3$ transition.

\begin{figure}
\begin{center}
  \includegraphics[scale = 0.50]{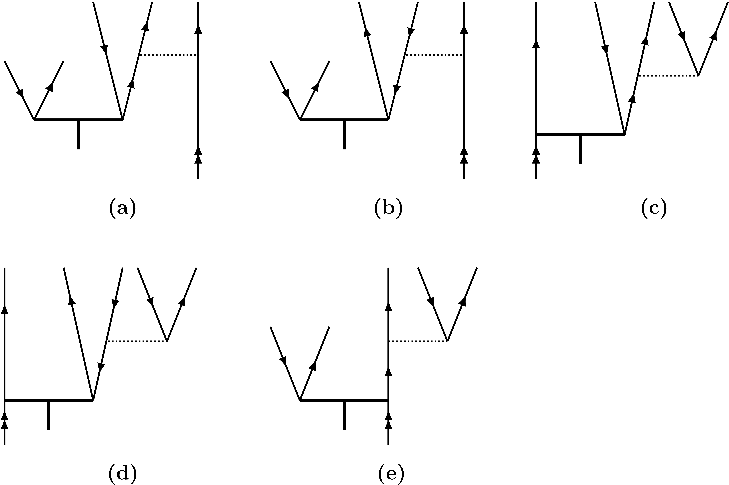}
  \caption{Perturbative ${\mathbf S}^{(1)}_3$ diagrams arising 
	from $g{\mathbf T}^{(1)}_2$ ((a) and (b)) and $g{\mathbf S}^{(1)}_2$ 
	((c), (d) and (e)) terms.}
  \label{fig_s3}
\end{center}
\end{figure}


\subsubsection{Breit and QED interactions}
In the Table \ref{nsd_tab} we have listed the contributions from Breit and QED 
corrections to ${E1}^{\rm NSD}_{\rm PNC}$. The QED results are the 
combined contribution from vacuum polarization and the self-energy 
correction. To compute the correction from vacuum polarization, we use the
Uehling potential \cite{uehling-35} with a modification to incorporate the 
finite size nuclear charge distribution \cite{fullerton-76,klarsfeld-77},
\begin{eqnarray}
	V_{\rm Ue}(r) &=& - \frac{2\alpha}{3r} \int^{\infty}_0 dx \ x \rho(x)
  \int^{\infty}_1 dt \sqrt{t^2 -1 } \left(\frac{1}{t^3} + \frac{1}{2t^5}\right)
                                     \nonumber \\
        && \times \left(e^{-2ct|(r-x)|} - e^{-2ct(r+x)}\right),
\end{eqnarray}
here, $\alpha$ is the fine structure constant, and should not be confused
with dipole polarizability and $\rho(x)$ is the finite size Fermi density 
distribution of the nuclear charge. The corrections from the self-energy to 
single-electron energies are incorporated through the model Lamb-shift operator 
introduced by Shabaev {\em et al.} \cite{shabaev-13} and using the code 
QEDMOD \cite{shabaev-15}. Quantitatively, the contributions with respect to 
PRCC values are shown Fig.\ref{fig_qed}(e) and Fig.\ref{fig_qed}(f). 
We observe two important trends in the contributions from the Breit and 
QED corrections. First, both contribute with the same phase for off-diagonal 
transitions, whereas they have opposite contributions for diagonal
transitions. Second, the contribution to the off-diagonal transition 
$4\rightarrow3$ is larger. In terms of magnitude, the largest cumulative 
contribution from Breit and QED corrections is $\approx$ 3.2\% of 
the PRCC value. This is significant and cannot be neglected.


\subsubsection{Perturbative triples}

To compute the contributions from perturbative triples to 
${E1}^{\rm NSD}_{\rm PNC}$, we used our approach for polarizability 
of one-valence systems \cite{ravi-22}. As discussed earlier, the PNC diagrams 
are topologically identical with the polarizability diagrams in the electronic 
sector. However, to obtain ${E1}^{\rm NSD}_{\rm PNC}$ we have to incorporate 
coupling with the nuclear spin. As the effect of external perturbation is 
included through the perturbed CC operators in PRCC theory, we pick the 
perturbative triples arising from $\mathbf{T}^{(1)}_2$ and 
$\mathbf{S}^{(1)}_2$. After contracting with the residual Coulomb interaction, 
$\contraction[0.4ex]{}{g}{}{T} g\mathbf{T}_2^{(1)}$ and 
$\contraction[0.4ex]{}{g}{}{S} g\mathbf{S}_2^{(1)}$, they lead to five 
${\mathbf S}^{(1)}_3$ diagrams shown in Fig. \ref{fig_s3}. A dominant 
contribution to ${E1}^{\rm NSD}_{\rm PNC}$ is expected from the contraction 
of $\mathbf{\tilde S}_3^{(1)}$ with $S^{(0)}_2$ due to larger amplitudes of 
$S^{(0)}$ for one-valence systems. This leads to 49 Goldstone diagrams 
\cite{ravi-22}, which are first calculated in the electronic sector and then 
coupled with the nuclear spin. The contributions from perturbative triples to 
${E1}^{\rm NSD}_{\rm PNC}$ amplitudes are listed in Table \ref{nsd_tab}. 
Consistent with the trend in $\alpha$, the contribution from triples to 
${E1}^{\rm NSD}_{\rm PNC}$ is small. The largest contribution to 
${E1}^{\rm NSD}_{\rm PNC}$ amplitudes is found to be $\approx$ 0.02\% .

\section{Theoretical uncertainty}

Considering the approximations employed in the computation of 
${E1}^{\rm NSD}_{\rm PNC}$, we identify four sources of theoretical 
uncertainties. The first one is the truncation of the orbital basis. From the 
convergence data of the ${E1}^{\rm NSD}_{\rm PNC}$ with basis size, for the 
basis up to $h-$symmetry, the change in ${E1}^{\rm NSD}_{\rm PNC}$ is 
$\approx 0.2\%$ when the optimal basis set with 169 orbitals is augmented.
This is noticeable from the plots in (Fig. \ref{fig_conv}(d)). To estimate the
uncertainty associated with the orbitals beyond $h$-symmetry we resort to 
the detailed analysis presented in Ref. \cite{ravi-21b}, our previous work on
$\alpha$ of one-valence atoms. Where we have shown that the largest overall 
contribution from the $i$, $j$ and $k$ symmetry orbitals to polarizability 
is $\approx 0.06$\%. Thus, the combined contribution from the orbitals 
of $k$ and higher symmetries is expected to be smaller and we take 0.1\% as 
an upper bound. So, the upper bound on the total uncertainty from the basis 
truncation is 0.3\%.  The second source of uncertainty is the truncation of the 
dressed operator $\bar{\mathbf{D}}$ to ${\mathbf D}+{\mathbf D}T^{(0)}
+{T^{(0)}}^\dagger {\mathbf D} + {T^{(0)}}^\dagger {\mathbf D}T^{(0)}$. 
To estimate the uncertainty from this source, we again use the findings from 
our previous work \cite{mani-10} where we showed that the combined 
contributions from terms with third-order in $T^{(0)}$ and higher is 
less than 0.1\%. So, we take this as the upper bound from this source.
The third source of uncertainty is the truncation of CC operators 
to singles and doubles. Among the higher excitations, triple excitations 
is the most prominent, and the dominant contribution is subsumed in 
the perturbative triples. So, to incorporate the dominant contribution 
from triple excitations, we have included the contributions from the 
perturbative triples arising from
$S{^{(0)\dagger}} \mathbf{D}g\mathbf{S_2^{(1)}}$ and
$S{^{(0)\dagger}} \mathbf{D}g\mathbf{T_2^{(1)}}$. The largest contribution 
from perturbative triples is observed to be $\approx 0.02\%$. Considering
this we take  0.1\% as an upper bound of the uncertainty due to the exclusion
of contributions non-perturbative triples and higher excitations. The last 
source of theoretical uncertainty is associated with the frequency-dependent 
Breit interaction which is not included in our calculations. To estimate an 
upper bound of this source we use the results in our previous
work \cite{chattopadhyay-14}, where using GRASP2K we estimated an upper 
bound of $0.13$\% for Ra. As Cs is lighter than Ra, the contribution is 
expected to be smaller. And, thus we take 0.13\% as the uncertainty from 
this source. There could be other sources of theoretical uncertainties, such 
as the higher order coupled perturbation of vacuum polarization and 
self-energy terms, etc. These, however, have much smaller contributions and 
their combined uncertainty could be below 0.1\%. Combining the upper bounds 
of all the four sources of uncertainties, we estimate a maximum theoretical 
uncertainty of 1\% in the computed values of ${E1}^{\rm NSD}_{\rm PNC}$ 
amplitudes.

\section{Conclusions}

We have implemented a Fock-space perturbed relativistic coupled-cluster
theory to compute the  NSD-PNC transition amplitudes. Using the method we
study the NSD-PNC of Cs atom. To check the wavefunctions obtained, we
calculated the excitation energies for low lying states, E1 transition
amplitudes, hyperfine structure constants, and dipole polarizability of
the ground and low lying excited states. And, to improve the accuracy
of the computed properties, the contributions from the Breit interaction
and QED corrections are incorporated.

Our results of excitation energies, E1 transition amplitudes and hyperfine
constants are in good agreement with the experimental data. Our results
of $\alpha$ using sum-over-states approach are marginally larger than
the experimental results. This may be attributed to the fact that the
sum-over-states approach does not include the contributions from all the
intermediate states. As can be expected, our results using FS-PRCC are in
excellent agreement with the experimental results. Our results of
$E1^{\rm NSD}_{\rm PNC}$ are in general lower than the previous calculations.
However, it should be noted that there is a variation among the previous
results.

From the quantitative analysis of the electron correlations, we find that
for the $\alpha$ of $6s_{1/2}$ and $7s_{1/2}$ states the dominant contribution
is from the dipolar mixing with $6p$ and $7p$ states, respectively.
For $6p_{1/2}$ and $7p_{1/2}$, however, the dominant dipolar
mixing is observed with $5d_{3/2}$ and $6d_{3/2}$ states, respectively.
Considering the $E1^{\rm NSD}_{\rm PNC}$, for both diagonal transitions
$3\rightarrow3$ and $4\rightarrow4$, the first two dominant PNC mixings
are from the $7p_{1/2}$ and $6p_{1/2}$ states.
Similarly, both the off-diagonal transitions $3\rightarrow4$ and
$4\rightarrow3$ are from the $6p_{1/2}$ and $7p_{1/2}$ states.
Based on our results, the largest combined contribution from the Breit and
QED corrections to $E1^{\rm NSD}_{\rm PNC}$ is $\approx$ 3.2\%. Hence it
is important to consider these corrections to obtain reliable value of
$E1^{\rm NSD}_{\rm PNC}$.

\begin{acknowledgments}
The authors wish to thank Palki Gakkhar for the useful discussions. Suraj Pandey 
acknowledges the funding support from the Ministry of Education, Govt. of India. 
BKM acknowledges the funding support from SERB, DST (CRG/2022/003845).
The results presented in the paper are based on the computations using 
High Performance Computing clusters Padum and TEJAS at the Indian Institute 
of Technology Delhi, New Delhi.
\end{acknowledgments}

\appendix 

\section{Convergence of excitation energies, E1 amplitudes and MHFS constants}

In Table \ref{basis_conv}, we show trend of the convergence of excitation 
energy, E1 transition amplitude, and magnetic dipole hyperfine structure 
constants as a function of the basis size. Similarly, the basis dependent 
results on $E1^{\rm NSD}_{\rm PNC}$ are provided in Table \ref{basis_e1pnc}. 
As it is evident from the table, 
all the properties converge to the order of $10^{-3}$ or less in the 
respective units of the properties.

\begin{table*}
  \caption{The excitation energies (cm$^{-1}$), E1 transition amplitudes (a.u.)
        and HFS constants (MHz) of Cs with increasing basis.}
\begin{ruledtabular}
\begin{tabular}{ccccccccccc}
        & \multicolumn{3}{c}{Excitation energies} & \multicolumn{4}{c}{MHFS} & \multicolumn{3}{c}{E1 amp.}\\
\hline
Orbitals/Basis          &
$6p_{1/2}$     &
$6p_{3/2}$     &
$7s_{1/2}$     &
$6s_{1/2}$     &
$6p_{1/2}$     &
$6p_{3/2}$     &
$7s_{1/2}$     &
$6p_{1/2} - 6s_{1/2}$ &
$6p_{1/2} - 7s_{1/2}$ &
$6p_{3/2} - 6s_{1/2}$  \\
\hline
        108(14s12p11d10f8g6h)   & 10896 & 11461 & 18156 & 2002 & 251 & 45 & 482 & 4.5888& 4.2909 & 6.4863\\
        119(15s13p12d11f9g7h)   & 10933 & 11499 & 18201 & 2060 & 260 & 46 & 495 & 4.5819& 4.2881 & 6.4747  \\
        130(16s14p13d12f10g8h)  & 10977 & 11544 & 18257 & 2110 & 266 & 48 & 506 & 4.5753& 4.2835 & 6.4616  \\
        141(17s15p14d13f11g9h)  & 11055 & 11629 & 18349 & 2173 & 273 & 50 & 519 & 4.5632& 4.2799 & 6.4395  \\
        152(18s16p15d14f12g10h) & 11130 & 11708 & 18443 & 2227 & 279 & 50 & 530 & 4.5491& 4.2733 & 6.4177  \\
        163(19s17p16d15f13g11h) & 11169 & 11749 & 18503 & 2263 & 284 & 51 & 536 & 4.5384& 4.2621 & 6.4039 \\
        173(21s19p18d15f13g11h) & 11169 & 11747 & 18504 & 2285 & 288 & 51 & 541 & 4.5377& 4.2607 & 6.4031     \\
        178(22s20p19d15f13g11h) & 11168 & 11746 & 18504 & 2292 & 289 & 51 & 543 & 4.5378& 4.2604 & 6.4032   \\
        183(23s21p20d15f13g11h) & 11168 & 11746 & 18503 & 2298 & 290 & 51 & 544 & 4.5378& 4.2603 & 6.4033  \\
        188(24s22p21d15f13g11h) & 11168 & 11746 & 18504 & 2302 & 290 & 51 & 545 & 4.5378& 4.2603 & 6.4033 \\
        193(25s23p22d15f13g11h) & 11168 & 11746 & 18504 & 2303 & 290 & 51 & 546 & 4.5378& 4.2603 & 6.4032  \\
	        \hline
\end{tabular}
\end{ruledtabular}
\label{basis_conv}
\end{table*}

\begin{table}
  \caption{The $E1^{\rm NSD}_{\rm PNC}$ amplitudes 
	(in the units of $iea_{0} \times 10^{-12} \mu^{'}_{W} $) 
	as a function of basis size.}
\begin{ruledtabular}
\begin{tabular}{ccc}
       Basis                 &
       $3 \rightarrow 3$     &
       $4 \rightarrow 3$   \\  
       \hline
       99(11s10p8d10f8g8h)   &1.8103& 5.0497 \\
       119(15s13p12d11f9g7h) &1.9137& 6.0420 \\
       129(17s15p14d11f9g7h) &1.9428& 6.2450 \\
       139(19s17p16d11f9g7h) &1.9683& 6.3496 \\
       144(20s18p17d11f9g7h) &1.9781& 6.3799 \\
       149(21s19p18d11f9g7h) &1.9844& 6.3973 \\
       154(22s20p19d11f9g7h) &1.9899& 6.4120 \\
       159(23s21p20d11f9g7h) &1.9926& 6.4223 \\
       164(24s22p21d11f9g7h) &1.9951& 6.4348 \\
       169(25s23p22d11f9g7h) &1.9959& 6.4416 \\
       174(26s24p23d11f9g7h) &1.9966& 6.4521 \\
\end{tabular}
\end{ruledtabular}
\label{basis_e1pnc}
\end{table}

\bibliography{references}

\end{document}